\documentclass[aps,twocolumn,prd,amssymb,eqsecnum,nofootinbib,floatfix,superscriptaddress]{revtex4}
\usepackage{epsfig}
\usepackage{amssymb}
\usepackage{amsmath}
\usepackage{amsfonts}
\usepackage{bm}
\newcommand{\be}{\begin{equation}}
\newcommand{\ee}{\end{equation}}
\newcommand{\bea}{\begin{eqnarray}}
\newcommand{\eea}{\end{eqnarray}}
\newcommand{\bes}{\begin{subequations}}
\newcommand{\ees}{\end{subequations}}

\newcommand{\nn}{\nonumber}

\begin{document}

\title{Gravitomagnetic resonant excitation of Rossby modes in
coalescing neutron star binaries}
\author{\'{E}anna \'{E}.\ Flanagan}
\affiliation{Center for Radiophysics and Space Research, Cornell
  University, Ithaca, New York, 14853}
\author{\'{E}tienne Racine}
\affiliation{Center for Radiophysics and Space Research, Cornell
  University, Ithaca, New York, 14853}
\affiliation{Theoretical Astrophysics, California Institute of Technology, Pasadena, California, 91125}

\begin{abstract}

In coalescing neutron star binaries, \textit{r}-modes in one of the
stars can be resonantly excited by the gravitomagnetic tidal field of
its companion.
This post-Newtonian gravitomagnetic driving of these modes
dominates over the Newtonian tidal driving previously
computed by Ho and Lai.
To leading order in the tidal expansion
parameter $R$/$r$ (where $R$ is the radius of the neutron star and $r$
is the orbital separation), only the $l=2, |m|= 1$ and $|m| = 2$
\textit{r}-modes are excited.
The tidal work done on the star through this
driving has an effect on the evolution of the
inspiral and on the phasing of the emitted gravitational wave signal.
For a neutron star of mass $M$, radius $R$, spin frequency $f_{\rm spin}$,
modeled as a $\Gamma =2$ polytrope,
with a companion also of mass $M$, the gravitational wave phase shift
for the $m=2$ mode
is $\sim 0.1 \,\text{radians}\,(R/10
\,\text{km})^4(M/1.4M_\odot)^{-10/3}(f_{\rm spin}/100
\,\text{Hz})^{2/3}$ for optimal spin orientation.
For canonical neutron star parameters this phase shift will likely not
be detectable by gravitational wave detectors such as LIGO, but if the
neutron star radius is larger it may be detectable
if the signal-to-noise ratio is moderately large.
The energy transfer is large enough to drive the mode into the
nonlinear regime if $f_{\rm spin} \agt 100 \, {\rm Hz}$.
For neutron star - black
hole binaries, the effect is smaller; the phase shift scales as
companion mass to the -4/3 power for large companion masses. The net
energy transfer from the orbit into the star is negative corresponding
to a slowing down of the inspiral. This occurs because the interaction
reduces the spin of the star, and occurs only for modes which
satisfy the Chandrasekhar-Friedman-Schutz instability criterion.

A large portion of the paper is devoted to developing a general
formalism to treat mode driving in rotating stars to post-Newtonian
order, which may be useful for other applications.  We also correct
some conceptual errors in the literature on the use of energy
conservation to deduce the effect of the mode driving on the
gravitational wave signal.

\end{abstract}
\maketitle

\section{Introduction and Summary}

\subsection{Background and motivation}
\label{sec:background}

One of the most promising sources for ground-based gravitational wave
observatories such as LIGO \cite{1992Sci...256..325A} and VIRGO
\cite{virgo} are coalescences of compact binary systems \cite{Cutler:2002me}.
The first searches for these systems have already been completed
\cite{Abbott:2003pj}, and future searches will be more sensitive.
For neutron star-neutron star (NS-NS) binaries, the most recent estimate of the
Galactic merger rate is $83^{+209}_{-66} \, {\rm Myr}^{-1}$
\cite{Kalogera:2003tn,Kalogera:2003tne}; extrapolating this to the
distant Universe using the method of Ref.\ \cite{Kalogera:2001dz}
yields a rate of $28^{+70}_{-22} \, {\rm yr}^{-1}$
within a distance of $200 \, {\rm Mpc}$.
The range of next-generation interferometers in LIGO is expected to be
$\sim 300 \, {\rm Mpc}$ \cite{Cutler:2002me}.
Coalescing binaries may well be the first detected sources of
gravitational waves.

\begin{figure}
\begin{center}
\epsfig{file=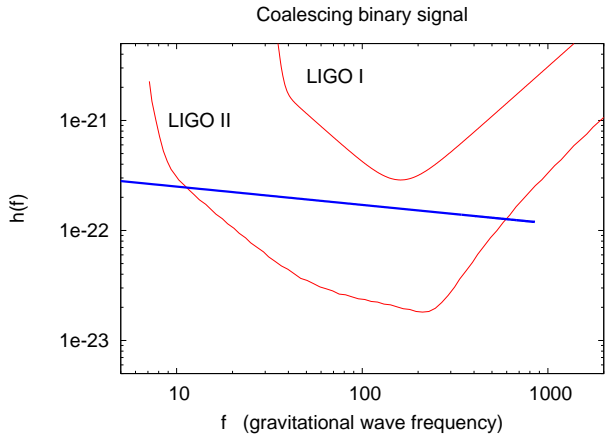,width=8.5cm}
\caption{Noise curves $h_{\rm rms}(f) = \sqrt{f S_h(f)}$ for initial and
next-generation LIGO interferometers \protect{\cite{noisecurveref}}
(shown in thin lines) together with the signal strength $h_c(f)$ for an
inspiralling binary of two $1.4 M_\odot$ neutron stars at a distance
of $200 \, {\rm Mpc}$ (thick line) \protect{\cite{hcdef}}.  The signal
will terminate near
the frequency of the innermost stable circular orbit, which is $\sim
850$ Hz as shown here if the neutron stars have radii of $10$ km and
are modeled as $\Gamma =2$ polytropes
\protect{\cite{Marronetti:2003hx}}.
Resonantly excited modes can give rise to small but potentially
measurable corrections to the phase of the
waveform in the early portion $10\,{\rm Hz} \lesssim f \lesssim 100 \, {\rm Hz}$
of the signal.}
\label{fig:noisecurves}
\end{center}
\end{figure}

The gravitational waveforms will carry a variety of different types of
information \cite{Cutler:2002me,1993PhRvL..70.2984C}:
First, in the early, low frequency phase of the signal,
the binary can be treated to a good approximation as consisting of two
spinning point masses.  The corresponding waveforms give detailed
information about the deviations of general relativity from Newtonian
gravity, and they have been computed to post-3.5-Newtonian
order \cite{Blanchet:2004ek}.
Second, toward the end of the inspiral (frequencies of order several
hundred Hertz, see Fig.\ \ref{fig:noisecurves}) the internal degrees of freedom
of the bodies start to appreciably influence the
signal, and there is the potential to infer information about the
nuclear equation of state.  This has prompted many numerical
simulations of the hydrodynamics of NS-NS mergers using various
approximations to general relativity (see, eg, Ref.\
\cite{Baumgarte:2002jm} and references therein).  Equation of state
information can also potentially be extracted from the waves energy
spectrum \cite{2002PhRvL..89w1102F}, and from the NS tidal disruption
signal for neutron star-black hole (NS-BH) binaries \cite{2000PhRvL..84.3519V}.

A third type of information potentially carried by the gravitational waves
is the effect of the internal structure of the bodies on the early,
low frequency ($10\, \text{Hz} \lesssim f \lesssim 100 \, \text{Hz}$) portion of
the signal, via the excitation of internal modes of
oscillation of one of the neutron stars by the tidal gravitational
field of its companion.  In this regime the
waveform's phase evolution is dominated by the point-mass dynamics,
and the perturbations due to the internal modes are small
corrections.  Nevertheless, if the accumulated phase shift due to the
perturbations becomes of order unity or larger, it could impede the
matched-filtering-based detection of NS-NS signals \cite{1993PhRvL..70.2984C}.
Alternatively the detection of a phase perturbation could give
information about the neutron star structure.  For these reasons the
excitation of neutron star modes has been studied in detail.

A given mode can be treated as a damped, driven harmonic
oscillator, described by the equation
\begin{equation}
{\ddot q} + \gamma {\dot q} + \omega_{\rm m}^2 q = A(t) \cos[ m \phi(t)].
\label{eq:eom00}
\end{equation}
Here $q$ is a generalized coordinate describing the mode, $\omega_{\rm
  m}$
is the mode frequency, $\gamma$ is a damping constant, $\phi(t)$ is
the orbital phase of the binary, $m$ is the azimuthal quantum number
of the mode, and $A(t)$ a slowly varying amplitude.
As the inspiral of the binary proceeds, the orbital frequency $\omega
= {\dot \phi}$ gradually increases with time.   Sufficiently early in
the inspiral, therefore, the mode will be driven below resonance:
$m \omega \ll \omega_{\rm m}$.
Because the inspiral is adiabatic we also
then have ${\dot A} / A \ll \omega_{\rm m}$ and ${\dot \omega} / \omega \ll
\omega_{\rm m}$.  Using these approximations Eq.\ (\ref{eq:eom00}) can be solved
analytically in this pre-resonance regime, and one finds that the
total energy absorbed by the mode up to time $t$ is
\be
E(t) = \frac{A(t)^2}{2 \omega_{\rm m}^2} + \gamma \int_{-\infty}^t dt'
\frac{\omega(t')^2 A(t')^2 }{ \omega_{\rm m}^4/m^2 + \omega(t')^2 \gamma^2}.
\label{eq:energyabsorbed}
\ee
This energy absorption causes the orbit to inspiral slightly faster
and changes the phase of the gravitational wave signal.

Mode excitations can have three different types of effects on the
gravitational wave signal:

\begin{itemize}

\item {\it Dissipative:} There is a phase perturbation due to the
second term in Eq.\ (\ref{eq:energyabsorbed}), corresponding to
energy dissipated by the mode.  This is a cumulative effect.  It is
dominated by the $l=2, m=\pm 2$ fundamental modes of the neutron star, and its
magnitude depends on the size of the damping constant $\gamma$, which
is determined by the fluid viscosity.  This effect was examined by
Bildsten and Cutler \cite{1992ApJ...400..175B}, who showed that the phase
perturbation is negligible for all physically reasonable values of the
viscosity.

\item {\it Adiabatic:} There is a phase perturbation due to the
first term in Eq.\
(\ref{eq:energyabsorbed}).  This corresponds to the instantaneous
energy present in the mode, as it adjusts adiabatically to sit at the
minimum of its slowly-evolving potential.  The perturbation to the
waveform is again dominated by the $l=2,m=\pm 2$ fundamental modes
\cite{1995MNRAS.275..301K}.
This effect has been studied analytically in Refs.\
\cite{1992ApJ...400..175B,1992ApJ...398..234K,1993ApJ...406L..63L,
1998PhRvD..58h4012T,Mora:2003wt} and numerically in Refs.\
\cite{1995MNRAS.275..301K,1994PThPh..91..871S,2001PhRvD..64j4007G,2002PhRvD..65j4021P,
2002PhRvD..66f4013B}.  It can also be inferred from sequences of 3D numerical
quasi-equilibrium models of NS-NS binaries \cite{
2002PhRvL..89w1102F,Marronetti:2003gk,Bejger:2004zx}.
The phase perturbation grows like $\omega(t)^{5/3}$ and
is of order a few cycles by the end of inspiral.  It may be marginally
detectable for some nearby events \cite{Hinderer:2005}.

\item {\it Resonant:} The frequencies of internal modes of a neutron star are typically
of order $\omega_{\rm m} \sim \sqrt{M / R^3}$, where $M$ and $R$ are the
stellar mass and radius.  (We use units with $G = c = 1$ throughout
this paper).  For a
NS-NS binary, this is of order the orbital
frequency at the end of the inspiral, $\sim 1000\, {\rm Hz}$.
Therefore, most modes are driven at frequencies below their natural
frequencies throughout the inspiral, and are never resonant.  However,
some mode do have frequencies $\omega_{\rm m}$ with $\omega_{\rm m} \ll
\sqrt{M/R^3}$, and these modes are resonantly excited as the gradually
increasing driving frequency $m \omega$ sweeps past $\omega_{\rm m}$.  During
resonance the effect of damping is negligible, and the width $\Delta
\omega$ of the resonance is determined by the inspiral timescale.
The energy deposited in the mode is thus enhanced compared to the
equilibrium energy [the first term in Eq.\ (\ref{eq:energyabsorbed})]
by the ratio of the inspiral timescale to the orbital period.  This
can be a significant enhancement.

\end{itemize}

Resonant driving of modes in NS-NS binaries has been studied
in Refs.\
\cite{1994ApJ...426..688R,1994MNRAS.270..611L,1999MNRAS.308..153H,
1995MNRAS.275..301K,2001PhRvD..64j4007G,
2002PhRvD..65j4021P,2002PhRvD..66f4013B},
and in the context of other binary systems in Refs.\
\cite{Willems:2002na,2003MNRAS.339...25R,Rathore:2004gs,Rathorethesis}.
One class of modes with suitably small frequencies are $g$-modes;
however the overlap integrals for these modes are so small that the
gravitational phase perturbation is small compared to unity
\cite{1995MNRAS.275..301K,1994ApJ...426..688R,1994MNRAS.270..611L}.
Another class are the $f$ and $p$-modes of rapidly rotating stars, where the
inertial-frame frequency $\omega_{\rm m}$ can be much smaller than the
corotating-frame frequency $\sim \sqrt{M/R^3}$.  Ho and Lai
\cite{1999MNRAS.308..153H} showed that the phase shifts due to these
modes could be large compared to unity.
However, the required NS spin frequencies are several hundred Hz,
which is thought to be unlikely in inspiralling NS-NS binaries.

A third class of modes are Rossby modes ($r$-modes)
\cite{1981A&A....94..126P,1982ApJ...256..717S}, for which the
restoring force is dominated by the Coriolis force.  For these modes the
mode frequency $\omega_{\rm m}$ is of order the spin frequency of the
star, and thus can be suitably small [$10 \, \text{Hz} \lesssim \omega_\text{
  m} / (2 \pi) \lesssim 100 \, \text{Hz}$].  Ho and Lai
\cite{1999MNRAS.308..153H} computed the Newtonian driving of these
modes, and showed that the phase shift is small compared to unity.

In this paper we compute the phase shift due to the post-Newtonian,
gravitomagnetic resonant driving of the $r$-modes.  Normally, post-Newtonian
effects are much smaller than Newtonian ones.  However, in this case
the post-Newtonian effect is larger.  The reason is that the Newtonian
driving acts via a coupling to the mass quadrupole moment
\begin{equation}
Q_{ij} = \int d^3 x \rho (x_i x_j - \frac{1}{3} \delta_{ij} {\bf x}^2)
\end{equation}
of the mode, where $\rho$ is the mass density.
For $r$ modes, this mass quadrupole moment vanishes to zeroth order in
the angular velocity $\Omega$ of the star; it is suppressed by a
factor $\sim \Omega^2 R^3 / M$.
By contrast, the gravitomagnetic driving acts via a coupling to the
current quadrupole moment
\begin{equation}
S_{ij} = \frac{1}{2} \int d^3 x \rho \left[ ({\bf x} \times {\bf
v})_i x_j + ({\bf x} \times {\bf v})_j x_i \right],
\end{equation}
where ${\bf v}$ is the velocity of the fluid.
This current quadrupole moment is nonvanishing to zeroth order in
$\Omega$, and the gravitomagnetic coupling is therefore enhanced over
the Newtonian tidal coupling by a factor of $\sim \Omega^2 R^3/M$.
This enhancement factor can exceed the usual factor $\sim v^2/c^2$ by
which post-Newtonian effects are suppressed, where $v$ is a velocity
scale.  It is for this same reason that the radiation-reaction
induced instability of $r$-modes in newly born neutron stars
\cite{1998ApJ...502..708A,1998ApJ...502..714F}
is dominated by the gravitomagnetic coupling and not the mass quadrupole
coupling.

\subsection{Effect of resonance on gravitational wave signal}

We now discuss the observational signature of a mode resonance in the
gravitational wave signal.  We denote by $\Phi(t)$ the phase of the
waveform, which is twice the orbital phase $\phi(t)$.  We denote by
$\Phi_{\rm pp}(t)$ the phase that one obtains from a point particle
model, neglecting the effects of the internal modes of the stars.
To the leading post-Newtonian order, this point-particle phase is
given from the quadrupole formula by the differential equation
\cite{Peters:1963ux}
\begin{equation}
{\dot \omega} = \frac{96}{5} \mu M_{\rm t}^{2/3} \omega^{11/3},
\label{eq:inspiral0}
\end{equation}
where $\omega = {\dot \phi}$ is the orbital angular velocity, and
$M_{\rm t}$
and $\mu$ are the total mass and reduced mass of the binary.

Consider now the phase $\Phi(t)$ including the effect of the mode.
We neglect the phase shift caused by the adiabatic response of
the mode at early times before the resonance [the first term in Eq.\
 (\ref{eq:energyabsorbed})], as this phase shift is small compared to
the effect of the resonance.  We also neglect the cumulative phase
shift due to damping.  Therefore at early times we have
$\Phi(t) = \Phi_{\rm pp}(t)$.  The duration of the resonance is short
compared to the inspiral timescale (see Sec.\ \ref{sec:estimates}
below).  After the resonance, the mode again has a negligible
influence on the orbit of the binary, and Eq.\ (\ref{eq:inspiral0})
applies once more.  However, the two constants of integration that
arise when solving Eq.\ (\ref{eq:inspiral0}) need not match between
the pre-resonance and post-resonance waveforms.  Therefore, the phase
can be written as\footnote{The sign of the $\Delta \Phi$ term in Eq.\
  (\protect{\ref{eq:signature}}) is chosen for later convenience.}
\begin{equation}
\Phi(t) = \left\{ \begin{array}{ll} \Phi_{\rm pp}(t) & \mbox{ $t - t_0
        \ll - t_{\rm res}$,}\\
        \Phi_{\rm pp}(t + \Delta t) - \Delta \Phi
& \mbox{ $t-t_0 \gg t_{\rm res}$.}\\ \end{array} \right.
\label{eq:signature}
\end{equation}
Here $t_0$ is the time at which resonance occurs, and $t_{\rm res}$ is
the duration of the resonance.\footnote{At intermediate times $|t-t_0|
  \sim t_{\rm res}$, the phase $\Phi(t)$
is not related in a simple way to $\Phi_{\rm pp}(t)$.  However, since the
duration of the resonance is typically a few tens of cycles (short
compared to the $\sim 10^4$ cycles of inspiral in the detector's frequency
band), the phase perturbation in this intermediate stage will not be
observable.}  The effect of the resonance is thus to
cause an overall phase shift $\Delta \Phi$, and also to cause a time
shift $\Delta t$ in the signal.  If energy is absorbed by the mode,
then the inspiral proceeds more quickly and $\Delta t$ is positive.

It turns out that the two parameters $\Delta t$ and $\Delta \Phi$
characterizing the effect of the resonance are not independent.
To a good approximation they are related by
\begin{equation}
\Delta \Phi = {\dot \Phi}_{\rm pp}(t_0) \Delta t,
\label{eq:constraint}
\end{equation}
as argued by Reisenegger and Goldreich \cite{1994ApJ...426..688R}, and
as derived in detail in Sec.\ \ref{sec:phasing} below.
Inserting this relation into Eq.\ (\ref{eq:signature}) and expanding to
linear order in the small parameter $\Delta t$ yields for the
late-time phase
\begin{equation}
\Phi(t) = \Phi_{\rm pp}(t) + \left[ {\dot \Phi}_{\rm pp}(t) - {\dot
    \Phi}_{\rm pp}(t_0) \right] \Delta t.
\label{eq:latetimephase}
\end{equation}
The physical meaning of the condition (\ref{eq:constraint}) is thus
that the resonance can be idealized as an instantaneous change in
frequency at $t=t_0$ with no corresponding instantaneous change in
phase.
Rewriting the phase (\ref{eq:latetimephase})
in terms of $\Delta \Phi$ using
(\ref{eq:constraint}) and using ${\dot \Phi}_{\rm pp} = 2 \omega$
gives
\begin{equation}
\Phi(t) = \Phi_{\rm pp}(t) + \left[ \frac{\omega(t)}{\omega_0} -1
     \right] \Delta \Phi,
\label{eq:signature1}
\end{equation}
where $\omega_0 = \omega(t_0)$ is the orbital frequency at resonance.
The phase perturbation due to the resonance therefore grows with time
after the resonance, and can become much larger than $\Delta \Phi$
when $\omega \gg \omega_0$.

One can alternatively think of the phase correction as being non-zero
before the resonance, and zero afterwards \cite{1994ApJ...426..688R}.
This is what would be
perceived if one matched a template with the portion of the observed
waveform after the resonance.  In other words, if we define
a point-particle waveform phase ${\tilde \Phi}_{\rm pp}(t)$ with different
conventions for the starting time and starting phase via
\begin{equation}
{\tilde \Phi}_{\rm pp}(t) \equiv \Phi_{\rm pp}(t + \Delta t) - \Delta
\Phi,
\end{equation}
then we obtain
\begin{equation}
\Phi(t) = \left\{ \begin{array}{ll}
{\tilde \Phi}_{\rm pp}(t) - \left[ \frac{\omega(t)}{\omega_0} -1
     \right] \Delta \Phi
 & \mbox{\ \  $t - t_0
        \ll - t_{\rm res}$,}\\
        {\tilde \Phi}_{\rm pp}(t)
& \mbox{\ \  $t-t_0 \gg t_{\rm res}$.}\\ \end{array} \right.
\label{eq:signature2}
\end{equation}
The phase correction $\Phi - {\tilde \Phi}_{\rm pp}$ is now a linear
function of frequency, interpolating between $\Delta \Phi$ at $\omega
=0$ to zero at resonance $\omega = \omega_0$.  Since this phase
correction never exceeds
$\Delta \Phi$, we see that the detectability criterion for the
resonance effect is $\Delta \Phi \gtrsim 1$.  Thus, the phase shifts $\gg
\Delta \Phi$ predicted by Eq.\ (\ref{eq:signature1}) are fictitious
in the sense that they are not measurable.

\subsection{Order of magnitude estimates}
\label{sec:estimates}

We now turn to an order of magnitude estimate of the mode excitation
and the phase shift $\Delta \Phi$, both for the Newtonian driving
studied previously and for the gravitomagnetic driving.

We consider a star of mass $M$ and radius $R$ with a companion of mass $M'$.
The orbital angular velocity at resonance is $\omega_0$; this is also
the mode frequency (up to a factor of the azimuthal quantum number $m$
which we neglect here).  The separation $r$ of the two stars at
resonance is then given by $\omega_0^2 \sim M_{\rm t}/r^3$, where
$M_{\rm t} = M + M'$ is the total mass.

Let the magnitude of the tidal acceleration be $a_{\rm tid}$, and let
the duration of the resonance be $t_{\rm res}$.  The mode can be
treated as a harmonic oscillator of mass $\sim M$ and frequency
$\omega_0$.  During the resonance,
the mode absorbs energy at the same rate as a free particle would, so
the total energy absorbed is
\be
\Delta E \sim M a_{\rm tid}^2 t_{\rm res}^2.
\label{eq:DeltaE}
\ee
Now the gravitational wave luminosity ${\dot E}_{\rm gw}$ is to a good
approximation unaffected by the resonance.  Therefore the loss of
energy $\Delta E$ from the orbit decreases the time taken to inspiral
by an amount $\Delta t = \Delta E / {\dot E}_{\rm gw}$.
From Eq.\ (\ref{eq:constraint}) the corresponding phase shift is
$\Delta \Phi \sim \Delta t/t_{\rm orb}$, where $t_{\rm orb} \sim
1/\omega_0$ is the orbital period.  This
gives
\be
\Delta \Phi \sim \frac{t_{\rm rr} \Delta E } {t_{\rm orb} E_{\rm orb}},
\label{eq:DeltaPhi11}
\ee
where $E_{\rm orb}$ is the orbital energy and $t_{\rm rr}$ is the
radiation reaction timescale.  This phase shift is just the energy
absorbed divided by the energy radiated per orbit.

Next, using $E_{\rm orb} \sim M M'/r$ and the formula (\ref{eq:DeltaE}) for
$\Delta E$ gives
\be
\Delta \Phi
\sim \frac{r}{M'} a_{\rm tid}^2 \frac{t_{\rm res}^2
  t_{\rm rr} }{ t_{\rm orb}}.
\ee
The resonance time $t_{\rm res}$ is the geometric mean of the orbital
and radiation reaction times $t_{\rm orb}$ and $t_{\rm rr}$.  This
follows from the fact that the orbital phase near resonance can be
expanded as
\be
\phi \sim \phi_0 + \omega_0 (t - t_0) + \frac{\omega_0}{t_{\rm rr}}
(t-t_0)^2 +
\label{eq:phaseexpand}
\ldots,
\ee
from the definition of the radiation reaction timescale $t_{\rm rr}$.
The mode is resonant when the quadratic term in
Eq.\ (\ref{eq:phaseexpand}) is small compared to unity, so that the
force on the mode is in phase with the mode's natural oscillation.
This gives $t_{\rm res} \sim \sqrt{t_{\rm orb} t_{\rm rr}}$, and
we get
\be
\Delta \Phi \sim \frac{r}{M'} a_{\rm tid}^2 t_{\rm rr}^2.
\ee
Finally, we use the scaling $t_{\rm rr} \sim r^4 / (\mu M_{\rm t}^2)$
for the radiation reaction time given by the
quadrupole formula, where $\mu$ is the reduced mass.
This gives
\be
\Delta \Phi \sim \frac{r^9 }{ M' \mu^2 M_{\rm t}^4} a_{\rm tid}^2.
\label{eq:phaseans0}
\ee

We now consider various different cases.  For the Newtonian driving of
a mode via a tidal field of multipole of order $l$, we have
\be
a_{\rm tid} \sim \frac{M' }{ r^2} \left({\frac{R}{r}} \right)^{l-1}.
\label{eq:atidal}
\ee
[The leading order, quadrupolar driving is $l=2$.]  Inserting this into
Eq.\ (\ref{eq:phaseans0}) and eliminating $r$ using $\omega_0^2 r^3
\sim M_{\rm t}$ gives
\be
\Delta \Phi \sim
\left( \frac{ M^3 }{\mu M_{\rm t}^2} \right) \left( \frac{R}{M}
\right)^5 \left( \frac{\omega_0^2}{M/R^3} \right)^{(l-5)/3},
\label{eq:phasefinal}
\ee
which agrees with previous analyses
\cite{1994ApJ...426..688R,1994MNRAS.270..611L} for $g$ and $f$ modes.

For the Newtonian driving of $r$-modes, the leading order driving is
$l=3$ rather than $l=2$ \cite{1999MNRAS.308..153H}.  Also the tidal
acceleration $a_{\rm tid}$ should be multiplied by the
suppression factor $\omega_0^2 R^3/M$ discussed in Sec.\
\ref{sec:background} above, and the final phase shift
(\ref{eq:phasefinal}) should be
multiplied by the square of this factor.  This gives
in the equal mass case $M = M'$
\be
\Delta \Phi \sim \frac{R^{10} \omega_0^{10/3}}{M^{20/3}}.
\ee
Now the frequency $m \omega_0$ of the dominant, $l=m=2$ $r$-mode is
related to the spin frequency $f_{\rm spin}$ of the neutron star
by $2 \omega_0 = 8 \pi f_{\rm spin}/3$, so we
obtain
\be
\Delta \Phi = k \frac{R^{10} f_{\rm spin}^{10/3}}{M^{20/3}},
\label{eq:rmodeNestimate}
\ee
where $k$ is a dimensionless constant of order unity.  Equation
(\ref{eq:rmodeNestimate}) agrees
with the results of
Ho and Lai \cite{1999MNRAS.308..153H}, who show that $k \approx 0.4$
for a $\Gamma=2$ polytrope, and for the $l=m=2$
$r$-mode.\footnote{See equation (4.13) of Ref.\ \cite{1999MNRAS.308..153H},
specialized to the equal mass case and maximized over the angle
$\beta$ between the spin of the star and the orbital angular momentum.}
This gives
\bea
&& \Delta \Phi_{\rm r-mode,N} = 8 \times 10^{-5} R_{10}^{10}
f_{s100}^{10/3} M_{1.4}^{-20/3},
\label{eq:HoLai}
\eea
where the subscript N denotes Newtonian and we have defined $R = 10
R_{10} \, {\rm km}$, $M = 1.4 M_{1.4} M_\odot$, and $f_{\rm spin} =
100 f_{s100} \, {\rm Hz}$.

For the gravitomagnetic driving of $r$-modes, the tidal acceleration is
given by
\be
a_{\rm tid} \sim \left( \frac{M' R}{r^3} \right) v_{\rm spin} v_{\rm orb}.
\ee
Here the first factor in brackets is tidal acceleration
(\ref{eq:atidal})
for Newtonian quadrupolar $l=2$ driving.  Since the interaction is
gravitomagnetic, it is suppressed by two powers of velocity relative
to the Newtonian interaction.  One power of velocity is associated
with the source of the gravitomagnetic field, and is the orbital
velocity $v_{\rm orb} \sim \omega_0 r$.  The second power of velocity
is that coming from the Lorentz-type ${\bf v} \times {\bf B}$ force
law, and is the internal fluid velocity in the neutron star due to its
spin, $v_{\rm spin} \sim \omega_0 R$.  Combining these estimates with
Eq.\ (\ref{eq:phaseans0}) and eliminating $\omega_0$ in favor of
$f_{\rm spin}$ gives
\be
\Delta \Phi \sim \frac{R^4 f_{\rm spin}^{2/3}}{M' M^2 M_{\rm t}^{1/3}}.
\label{eq:ans12}
\ee
In the equal mass case $M = M'$ this gives
\bea
&& \Delta \Phi_{\rm r-mode,PN} \sim 3.4 \
R_{10}^{4} f_{s100}^{2/3} M_{1.4}^{-10/3}.
\label{eq:ans2}
\eea
Here the subscript PN denotes post-Newtonian.

\subsection{Results and implications}

\begin{figure}
\begin{center}
\epsfig{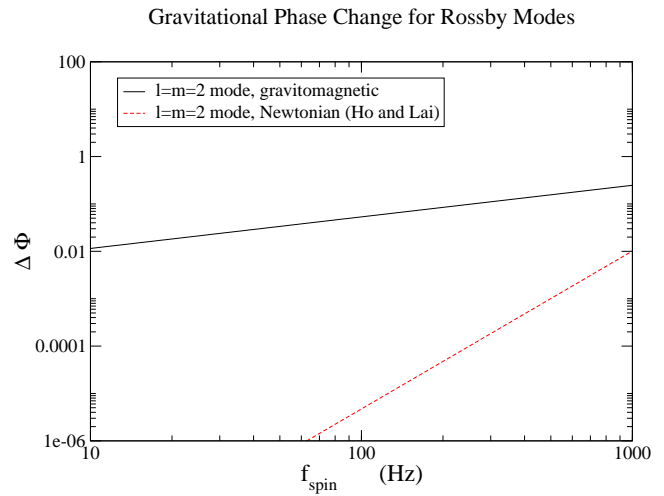}
\caption{The magnitude of the gravitational-wave phase shift $\Delta
\Phi$ due to the $r$-mode
resonance as a function of the spin frequency $f_{\rm spin}$ of the
neutron star, for both the Newtonian and gravitomagnetic driving, for
a binary with masses $M = M' = 1.4 M_\odot$, and for a neutron star
radius of $R = 10 \, {\rm km}$.  The star is modeled as a $\Gamma=2$ polytrope.
The mode plotted is $l=2$, $m=2$ in each case.  The angle $\psi$
between the spin and the orbital angular
momentum has been chosen in each case to maximize the phase shift;
$\psi$ is $34.4^\circ$ in the Newtonian case and $60^\circ$ in the
post-Newtonian case.
The resonance occurs at an orbital frequency of
$4 f_{\rm spin} / 9$ in the Newtonian case and at $4 f_{\rm spin}/3$
in the post-Newtonian case. The corresponding gravitational wave
frequencies are $8 f_{\rm spin} / 9$ and $8 f_{\rm spin}/3$
respectively. It can be seen that the gravitomagnetic driving
dominates.}
\label{fig:result}
\end{center}
\end{figure}

The fact that the order-of-magnitude estimate (\ref{eq:ans2}) of the
phase shift is of order unity, and thus potentially detectable
in the gravitational wave signal via matched
filtering \cite{1993PhRvL..70.2984C}, is the primary motivation for
the computation of
this paper.  We compute the dimensionless coefficient in Eq.\
(\ref{eq:ans12}), and find for the case of a $\Gamma=2$ polytrope for
the $l=2$, $m=2$ mode [cf.\ Eq.\ (\ref{eq:ans2ssa}) below, multiplied
by two to convert from orbital phase shift $\Delta \phi$ to
gravitational wave phase shift $\Delta \Phi$]
\be
\Delta \Phi_{\rm r-mode,PN} = -0.06 \ R_{10}^4 f_{s100}^{2/3}
M_{1.4}^{-10/3}.
\label{eq:ans2ss}
\ee
This is somewhat smaller than the estimate (\ref{eq:ans2}).
Figure \ref{fig:result} compares this
result to the Newtonian phase shift (\ref{eq:HoLai}).
In Table \ref{table:1} we show the phase shift $\Delta \Phi_{22} = 2
\Delta \phi_{22}$ that will occur at coalescence for the five known
double neutron star binaries.\footnote{The orbital motion
of these five systems is eccentric today, but radiation reaction will have
  circularized the orbits by the time resonance occurs. It is thus
  consistent to use our circular-orbit results for these binaries.}
For B1913+16 we used the known value $\psi=22^\circ$ of the
angle between the pulsar spin and the orbital angular
momentum \cite{1998ApJ...509..856K}, and for the remaining binaries for
which $\psi$ is unknown we took an average of the angular
factors appearing in the phase lag (\ref{eq:deltaphi22ans}).  We
assumed a $\Gamma =2$ polytrope equation of state and a
radius of 10 km for all pulsars.

\begin{table}
\begin{tabular}{|c|c|c|c|c|}
\hline
System & spin period& $M$ & $M^\prime$ & phase lag $\Delta \Phi_{22}$ \\
\hline
B1913+16 &59.0&1.441&1.387& -0.005 \\
B1534+12 &37.9&1.339&1.339& -0.009 \\
B2127+11C &30.5&1.358&1.354& -0.010 \\
J0737-3039 & 22.7&1.34&1.25& -0.013\\
J1756-2251 & 28.5&1.40&1.18& -0.011\\
\hline
\end{tabular}
\caption{List of gravitational wave phase lag parameters
that will occur for the
$l=2,m=2$ resonance for
the known double neutron star binaries when they coalesce.
For B1913+16, the known
value for the misalignment angle $\psi$ ($22^\circ$) was
used; for the others, the average value of the angular
factors was used. The pulsar mass $M$ and companion mass
$M^\prime$ are in solar masses and the pulsar spin period is in
milliseconds. The parameter values were taken from
Ref.\ \cite{2005LRR.....8....7L}.}
\label{table:1}
\end{table}

The phase shift (\ref{eq:ans2ss}) is sufficiently small that it is
unlikely to be detectable in the gravitational wave signal for most
detected binary inspirals.  Even for the most favorable conceivable
neutron star parameters of $M = 1.0 \, M_\odot$ and $R = 15 \, {\rm km}$, the
phase shift is $\sim 0.8$ radians.  An analysis of the detectability
of mode resonances indicates that the minimum detectable value of
$\Delta \Phi$ is $\Delta \Phi \sim 2$ for $f_{\rm spin} = 100 \, {\rm
Hz}$ at a signal to noise ratio of $S/N = 10$ \cite{Balachandran:2005}.
Therefore the minimum signal to noise ratio necessary for detection of
the effect is between $\sim 30$ and $\sim 300$, depending on the
neutron star parameters.  It is conceivable that the effect could be
detected in some of the closest detected binary inspirals with
advanced interferometers.

One of the interesting features of the interaction is that the phase
shift is negative, corresponding to energy being transferred from the
star to the orbit rather than the other way around.  This arises
because the star is rotating and the interaction reduces the spin of
the star, tapping into its rotational kinetic energy.\footnote{This
follows from the fact that at fixed angular momentum, the energy of a
star is minimized at uniform rotation.  Therefore the only way the
total energy of a rotating star can be reduced is via a reduction of the
spin.}  In Sec.\ \ref{sec:evol2} below we show
that this sign of energy
transfer is only possible for modes which satisfy the
Chandrasekhar-Friedman-Schutz mode instability criterion
\cite{1970PhRvL..24..762C,1978ApJ...222..281F} of being
retrograde in the corotating frame and prograde in the
inertial frame.

Although the $r$-mode resonance may not be detectable, it is
nevertheless probably the strongest of all the mode resonances in the
early phase of the gravitational wave signal.  The energy transfer
corresponding to the phase shift (\ref{eq:ans2ss}) is from Eq.\
(\ref{eq:DeltaPhi11})
\be
\Delta E \sim (1.4 \times 10^{48} \, {\rm erg})\, R_{10}^4 f_{s100}^3,
\ee
which is fairly large.  The
corresponding dimensionless
amplitude $\alpha$ of the $r$-mode in the notation of Ref.\
\cite{prl...80...4843} is
\be
\alpha \sim 0.3 \, R_{10}^2 f_{s100} M_{1.4}^{-1}.
\ee
This amplitude is sufficiently large that nonlinear mode-mode
interactions should be important, as in the case of unstable $r$-modes
in newly born neutron stars and in low mass X-ray binaries
\cite{prl...80...4843}.  In those contexts the mode-mode coupling
causes the mode growth to saturate at $\alpha \sim 10^{-4}$
\cite{Schenk:2001zm,Morsink:2002ut,Arras:2002dw,Brink:2004bg,Brink:2004qf,Brink:2004kt}.
Here however the $r$-mode is driven on much shorter timescales, and the
mode-mode coupling likely does not have time to operate efficiently.
It is possible that the type of nonlinear effects studied numerically in Refs.\
\cite{Gressman:2002zy,Lindblom:2001sg} (generation of differential
rotation, shocks, wave breaking) could occur.

\subsection{Organization of this paper}

The remainder of this paper is organized as follows.  In Secs.\
\ref{sec:prelim}, \ref{sec:modeevolution} and \ref{sec:phasing}
we analyze resonant driving of modes in Newtonian
gravity, building on the previous studies of Refs.\
\cite{1994ApJ...426..688R,1994MNRAS.270..611L,1999MNRAS.308..153H,Rathore:2004gs,Rathorethesis}.
The time delay parameter $\Delta t$ can be derived using energy
conservation which we discuss in Appendix \ref{sec:energycons}, but
the relation (\ref{eq:constraint}) between the phase shift parameter $\Delta \Phi$
and $\Delta t$ requires an analysis of the orbital equations of
motion, which we give in Sec.\ \ref{sec:phasing}.
In Sec.\ \ref{sec:rmodedriving} and Appendix \ref{app:Y}
we show that the gravitomagnetic
driving of modes can be computed to leading order using Newtonian
stellar perturbation theory supplemented by a gravitomagnetic external
acceleration, and we deduce the phase shifts generated by the driving
of $r$-modes.

\section{Newtonian mode-orbit coupling}
\label{sec:prelim}

In this section we describe a model dynamical system that describes
the Newtonian excitation of modes in inspiralling binary systems.  We
generalize previous treatments \cite{Rathore:2004gs,Rathorethesis}
to allow for stellar rotation.
The dynamical system will actually be general
enough to encompass the post-Newtonian driving of $r$-modes, as we
discuss in Sec.\ \ref{sec:rmodedriving} below.

\subsection{Mode expansions in rotating Newtonian stars}
\label{sec:modeexpand}

We start by reviewing the evolution equations for mode amplitudes in rotating
Newtonian stars, following the treatment in Ref. \cite{Schenk:2001zm}.
We assume that the background star is uniformly rotating with angular
velocity ${\bf \Omega}$ and mass $M$.  We denote by $(\bm{x},t)$ the
inertial frame coordinates and by $(\bar{{\bm x}},t)$ the
co-rotating frame coordinates.  These coordinate systems are oriented
so that ${\bf \Omega}$ is in the $z$ direction.
Below we will use spherical
polar coordinates $(r,\theta_s,\bar{\phi_s})$ for $\bar{{\bm x}}$; the
corresponding inertial frame coordinates are $(r,\theta_s,\phi_s)$
where
\be
{\phi_s} = \bar{\phi}_s + \Omega t.
\label{anglerelation}
\ee

A mode of a rotating star is a pair $(\bm{\xi}(\bar{{\bm x}}), \omega)$
of a mode function $\bm{\xi}(\bar{{\bm x}})$ and a rotating-frame
frequency $\omega$, such that the rotating-frame Lagrangian
displacement given by
\be
\bm{\xi}(\bar{{\bm x}},t) = \bm{\xi}(\bar{{\bm x}}) e^{- i \omega t}
\ee
is a solution to the linearized hydrodynamic equations.
We label the different modes with $\omega \leq 0$
\footnote{
In a rotating star, to obtain a complete set of modes for the phase
space mode expansion, one can either choose all the modes with $\omega
\ge0$ or all the modes with $\omega \le 0$.  Here we choose the
latter (so the relevant {\it r}-modes have $m\ge 0$); the opposite
convention was used in Ref.\ \protect{\cite{Schenk:2001zm}}.} by an index
$j$; the
$j$th mode is $(\bm{\xi}_j(\bar{{\bm x}}), \omega_{j})$.
It can then be shown that the Lagrangian displacement $\bm{\xi}(\bar{{\bm
x}},t)$ and its time derivative ${\dot {\bm{\xi}}}(\bar{{\bm x}},t)$ can be
expanded in the phase-space mode expansion \cite{Schenk:2001zm}
\be
\left[\begin{array}{c} \bm{\xi} \\ \dot{\bm{\xi}} \end{array}\right] =
\sum_{j}c_{j}(t)\left[\begin{array}{c} \bm{\xi}_{j} \\
    -i\omega_{j}\bm{\xi}_{j} \end{array}\right] +
c^\ast_{j}(t)\left[\begin{array}{c} \bm{\xi}^\ast_{j} \\
    i\omega_{j}\bm{\xi}^\ast_{j} \end{array}\right].
\label{eq:modeexpansion}
\ee
In rotating stars, one cannot in general find a basis of modes which
are real, so the expansion coefficients $c_j(t)$ are complex in
general.

If there is an external acceleration
$\bm{a}_{\text{ext}}(\bar{{\bm x}},t)$ acting on the star, then the equation
of motion for the $j$th mode is \cite{Schenk:2001zm}
\be\label{dotxrmode}
\dot{c}_j + i\omega_{j} c_j = \frac{i}{b_j}\langle \bm{\xi}_j , \bm{a}_{\text{ext}} \rangle,
\ee
where the inner product is defined by
\be\label{modedrivingterm}
\langle \bm{\xi}_j , \bm{a}_{\text{ext}} \rangle = \int d^3\bar{x}\,\rho\,
\bm{\xi}_j^\ast \cdot \bm{a}_{\text{ext}}
\ee
and $\rho(\bar{{\bm x}})$ is the background density.  The constant $b_j$ is
given by
\be
b_j = \langle \bm{\xi}_j , 2 \omega_j \bm{\xi}_j + 2 i {\bf \Omega}
\times \bm{\xi}_j \rangle.
\label{eq:bdef}
\ee
We normalize the modes using the convention
\be\label{modenorm}
\langle \bm{\xi}_j , \bm{\xi}_j\rangle =
M R^2,
\ee
where $R$ is the stellar radius.

An equivalent second-order form of the equation of motion can be
obtained by
differentiating Eq.\ (\ref{dotxrmode}) and
eliminating ${\dot c}_j$ in the result using Eq.\ (\ref{dotxrmode}).
This gives
$${\ddot c}_j + \omega_{j}^2 c_j = {\dot f}_j - i \omega_j f_j,
$$
where
$
f_j(t) \equiv i\langle \bm{\xi}_j, \bm{a}_{\text{ext}} \rangle / b_j.
$
This is the form
of the equation of motion discussed in the Introduction,
cf.\ Eq.\  (\ref{eq:eom00}).

\subsection{Mode-orbit coupling equations}

We now consider a binary system of a star of mass $M$, whose modes are
excited, and a companion of mass $M'$ which we treat as a point mass.
We denote the total mass $M + M'$ by $M_{\rm t}$, and the reduced
mass $M M'/M_{\rm t}$ by $\mu$.
We denote by ${\bm x} = (r,\theta,\phi)$ the displacement of $M'$ with
respect to $M$ in polar coordinates.  We orient these polar
coordinates so that the motion is confined to the equatorial plane
$\theta=\pi/2$.
The equations of motion describing this system are
\bes\label{system}
\bea\label{dotx}
&& \dot{c}_j + i\omega_j c_j = f_j, \\\label{dotpr}
&& {\ddot r} - r {\dot \phi}^2 + \frac{M_{\rm t}}{r^2} = a_{\hat{r}}^{\text{modes}} + a_{\hat{r}}^{\text{diss}},
\\\label{dotpphi}
&& r {\ddot \phi} + 2 {\dot r} {\dot \phi}
= a_{\hat{\phi}}^{\text{modes}} + a_{\hat{\phi}}^{\text{diss}}.
\eea
\ees
Here $c_j$ and $\omega_j$ are the amplitude and rotating-frame
frequency of the $j$th mode.  The right hand sides of Eqs.\
(\ref{dotpr}) and (\ref{dotpphi})
are the components of the total relative acceleration
${\bm a}^{\text{modes}} + {\bm a}^{\text{diss}}$
on the orthornomal basis ${\bm e}_{\hat r}$, ${\bm e}_{\hat \phi}$, where
$\bm{a}^{\text{modes}}$ describes the back-reaction of excited modes
onto orbital motion and $\bm{a}^{\text{diss}}$ is the back-reaction of
emission of gravitational radiation onto orbital motion.
The latter acceleration term is obtained
from the Burke-Thorne radiation-reaction acceleration \cite{mtw}
\begin{equation}
a_i^{\rm diss}(t,{\bf x}) = - \frac{2}{5} \frac{d^5 Q^{\rm TF}_{ij} }{dt^5}(t) x_j,
\end{equation}
where $Q_{ij}$ is the quadrupole moment of the binary and TF means
trace-free part.
In evaluating this expression we include only the contribution
$Q_{ij} = \mu x_i x_j$ to
the quadrupole from the orbital degrees of freedom, neglecting the
contribution from the modes.  To evaluate the time derivatives of
$Q_{ij}$ we use
the Newtonian orbital equations of motion with the mode coupling terms
omitted.   We also neglect the effect of
gravitational wave damping on the modes themselves.
This yields the components of the dissipative acceleration
\be
a^{\rm diss}_{\hat r} = \frac{16 \mu M_{\rm t}}{5 r^3} {\dot r} \left[
{\dot r}^2 + 6 r^2 {\dot \phi}^2 + \frac{4 M_{\rm t}}{3 r} \right]
\label{eq:adissr}
\ee
and
\be
a^{\rm diss}_{\hat \phi} = \frac{8 \mu M_{\rm t}}{5 r^2} {\dot \phi} \left[
9 {\dot r}^2 - 6 r^2 {\dot \phi}^2 + \frac{2 M_{\rm t}}{r} \right].
\label{eq:adissphi}
\ee

The function $f_j$ in Eq.\ (\ref{dotx}) describes the coupling
between mode $j$ and the tidal gravitational field of the companion
object.  For Newtonian mode-orbit coupling, it can be derived as
follows \cite{1999MNRAS.308..153H}.  Insert
into the right hand side of Eq.\ (\ref{dotxrmode})
the acceleration ${\bm a}_{\rm ext} = - {\bm \nabla} \Phi_{\rm ext}$,
where the external potential is
\bea
\Phi_{\rm ext} &=& - \bm{\nabla} \frac{M'}{| {\bm x} - {\bm x}(t)| }
\nn \\
&=& - M' \sum_{l'=2}^\infty \sum_{m'=-l'}^{l'} W_{l'm'} \frac{r^l e^{- i m'
    \phi(t)} }{ r(t)^{l+1}}
Y_{l'm'}(\theta,\phi). \nonumber \\
\label{phiext}
\eea
Here the coefficient $W_{l'm'}$ is given by Eq.\ (2.2) of Ref.\ \cite{1999MNRAS.308..153H},
and ${\bm x}(t) = (r(t),\pi/2,\phi(t))$ is the location of the companion.
The spherical harmonics that appear in Eq.\ (\ref{phiext}) are related to
to spherical harmonics of the inertial-frame coordinates
$(r,\theta_s,\phi_s)$ which are aligned with the spin axis of the star by
\be
Y_{l'm'}(\theta,\phi) = \sum_m {\cal D}^{(l')}_{mm'}
  Y_{l'm}(\theta_s,\phi_s),
\ee
where ${\cal D}^{(l')}_{mm'}$ is the Wigner ${\cal D}$-function
\cite{1999MNRAS.308..153H}.
Finally use the relation (\ref{anglerelation}) between the inertial
and corotating frame coordinates $\phi_s$ and ${\bar \phi}_s$.
The result is
\be
f_j = \sum_{l'm'} Q_{j,l'm'} \, \frac{ e^{-
    i m' \phi(t) + i m_j \Omega t} }{r(t)^{l'+1}},
\label{fjans}
\ee
where $m_j$ is the azumithal quantum number of the mode $\bm{\xi}_j$,
and
\be
Q_{j,l'm'} = \frac{i M'}{b_j} W_{l'm'} {\cal D}^{(l')}_{m_jm'}
\left< \bm{\xi}_j , \bm{\nabla} [r^{l'} Y_{l'm_j}(\theta_s,{\bar \phi}_s)]\right>.
\ee
Note that there are two different azimuthal quantum numbers in the
final expression (\ref{fjans}), the quantum number $m_j$ of the mode
which controls the number of factors of the phase $e^{i \Omega t}$,
and the quantum number $m'$ which controls the number of factors of
the phase $e^{- i \phi(t)}$.  These two quantum numbers
coincide when the spin axis of the star is aligned with the orbital
angular momentum, but not in general.

For the Newtonian case, in the absence of dissipation, the dynamical
system (\ref{system}) can be derived from the
Hamiltonian
\be
H = \frac{p_r^2 }{ 2 \mu} + \frac{p_\phi^2}{2 \mu r^2} - \frac{\mu
  M_t}{r} + \sum_j b_j
\left[ \omega_j |c_j|^2 + i f_j c_j^* - i f_j^* c_j \right],
\label{eq:Hamiltonian}
\ee
where $p_r = \mu {\dot r}$, $p_\phi =\mu r^2 {\dot \phi}$,
together with the Poisson brackets $\{ c_j, c_k \}=0$, $\{ c_j,
c_k^*\} = - i \delta_{jk}/b_j$.
This allows us to read off the orbital acceleration
$\bm{a}^{\text{modes}}$ due to the modes:
\bes
\label{eq:amodesformula}
\bea
\label{eq:arhat}
a_{\hat r}^{\text{modes}} &=&\frac{i}{\mu} \sum_j b_j \left(c_j \frac{\partial
  f_j^*}{\partial r} - c_j^* \frac{ \partial f_j}{\partial r}\right),  \\
a_{\hat \phi}^{\text{modes}} &=& \frac{i}{\mu r} \sum_j b_j \left(c_j \frac{\partial
  f_j^*}{\partial \phi} - c_j^* \frac{ \partial f_j}{\partial
  \phi}\right).
\label{eq:aphihat}
\eea
\ees

From the formula (\ref{fjans}) we see that a given mode $j$ is driven
by a sum of terms labeled by $(l',m')$, the effects of which superpose
linearly.  For the remainder of this section and in
Secs. \ref{sec:modeevolution} and \ref{sec:phasing} we will
focus attention on a particular mode $j$ and on a particular driving
term $(l',m')$.  Using the notations $A_j(r) =
|Q_{j,l'm'}|/r^{l'+1}$ and $e^{i \chi_j} = Q_{j,l'm'}/|Q_{j,l'm'}|$,
the coupling function (\ref{fjans}) and the orbital
acceleration (\ref{eq:aphihat}) can be written as
\bes
\label{eq:Newtdriving}
\bea
\label{eq:fjformula}
f_j &=& A_j[r(t)] \, e^{- i m' \phi(t) + i m_j \Omega t + i \chi_j},\\
a_{\hat \phi}^{\text{modes}} &=& - \frac{2 m' b_j A_j[r(t)]}{\mu r(t)}
\Re \left[ c_j^* e^{- i m' \phi(t) + i m_j \Omega t + i \chi_j}
  \right]. \nn \\
\label{eq:ahatphi1}
\eea
\ees
Although these formulae (\ref{eq:Newtdriving}) were derived for
Newtonian tidal driving, we shall see in Sec.\ \ref{sec:rmodedriving}
below that they also apply to gravitomagnetic driving in the vicinity
of a resonance, with appropriate choices of the functions $A_j(r)$ and
phases $\chi_j$.

\subsection{New orbital variables}

In order to analyze the effects of the mode coupling on the orbital
motion it is useful to make a change of variables, from
$r(t)$, $\phi(t)$, ${\dot r}(t)$, ${\dot \phi}(t)$ to new variables
$p(t)$, $\phi(t)$, $e(t)$ and $\phi_0(t)$ that characterize the Newtonian
ellipse that is instantaneously tangent to the orbit.
Here $p$ is the instantaneous semilatus
rectum and $e$ is the instantaneous eccentricity.  These variables are
defined by the equations
\bes\label{eq:newvars}
\bea
\label{d}
r &=& \frac{p}{1 + e c_\phi}, \\
{\dot r} &=& \sqrt{\frac{M_{\rm t}}{p}} e s_\phi, \\
{\dot \phi}&=& \sqrt{\frac{M_{\rm t}}{p^3}} \left[ 1 + e c_\phi \right]^2,
\eea
\ees
where $c_\phi = \cos(\phi - \phi_0)$ and $s_\phi = \sin(\phi - \phi_0)$.
In terms of these variables the orbital equations become
\bes
\label{eq:newvars1}
\bea
\label{eq:pdot00}
{\dot p} &=& \frac{ 2 p^{3/2}}{\sqrt{M_{\rm t}} (1 + e c_\phi) }
a_{\hat \phi}\\
\label{eq:phidot00}
{\dot \phi} &=& \frac{ \sqrt{M_{\rm t}}}{p^{3/2}} (1 + e c_\phi)^2,
 \\
\label{eq:edot00}
{\dot e} &=& \frac{ \sqrt{p}}{ 2 \sqrt{M_{\rm t}} (1 + e c_\phi)}   \nn \\
&&\times  \bigg[ (3 e + 4 c_\phi + e c_{2\phi}) a_{\hat \phi} + (2 s_\phi + e s_{2\phi}) a_{\hat r} \bigg], \nn \\
&& \, \\
{\dot \phi}_0 &=& \frac{ - \sqrt{p}}{ 2 e \sqrt{M_{\rm t}} (1 + e c_\phi)}  \nn \\
&& \times \bigg[ ( e + 2 c_\phi + e c_{2\phi}) a_{\hat r} - (4 s_\phi + e s_{2\phi}) a_{\hat \phi} \bigg]. \nn \\
\label{eq:phi0dot00}
\eea
\ees
Here $c_{2\phi} = \cos[2 (\phi - \phi_0)]$, $s_{2\phi} = \sin[2 (\phi
  - \phi_0)]$, and $a_{\hat r} = a_{\hat{r}}^{\text{modes}} +
a_{\hat{r}}^{\text{diss}}$ and $a_{\hat \phi} =
a_{\hat{\phi}}^{\text{modes}} + a_{\hat{\phi}}^{\text{diss}}$ are the
components
of the total acceleration in the radial and tangential directions.

\subsection{The circular approximation}

In this paper we restrict attention to situations where the
eccentricity $e$ is negligibly small before, during and after the mode
resonance.
We will justify this assumption in Sec.\ \ref{sec:circularcheck}
below by estimating the magnitude of the eccentricity generated during the resonance.
Using this approximation we can simplify the set of equations as follows.  We can
drop Eqs.\ (\ref{eq:edot00}) and (\ref{eq:phi0dot00}), and we simplify the
remaining equations (\ref{dotx}), (\ref{eq:pdot00}),
and (\ref{eq:phidot00}) using $e=0$.  Using the expressions (\ref{eq:adissr})
and (\ref{eq:adissphi}) for the dissipative components of the
acceleration, and the mode-orbit coupling terms (\ref{eq:Newtdriving}), now gives
\bes\label{nicesystem2}
\bea\label{nicedotx2}
 \dot{c}_j + i\omega_j c &=&
A_j(p) \, e^{- i m' \phi(t) + i m_j \Omega t + i \chi_j},
 \\\label{nicedotpr2}
 {\dot p}+ \frac{64 \mu M_{\rm t}^2}{5 p^3} &=&
- \frac{4 m' b_j A_j(p) \sqrt{p}}{\mu \sqrt{M_{\rm t}}}
\nn\\
&&\times \Re \left[ c_j^* e^{- i m' \phi(t) + i m_j \Omega t + i \chi_j}
  \right], \ \ \ \   \\
\label{nicedotpphi2}
 {\dot \phi} - \frac{ \sqrt{M_{\rm t}}}{p^{3/2}} &=&0.
\eea
\ees
Note that only the $\phi$ component of the acceleration contributes in
this approximation.

\subsection{The no-backreaction approximation}

When the amplitude of the mode coupling $A_j$ is small we can solve the system
(\ref{nicesystem2}) of equations by neglecting the backreaction of the
mode excitation on the orbital motion.
More precisely, we
(i) solve the orbital equations
(\ref{nicedotpr2}) and (\ref{nicedotpphi2}) for the zeroth order
solutions $p(t)$ and $\phi(t)$, neglecting the mode coupling;  (ii)
insert those zeroth order solutions into Eq.\ (\ref{nicedotx2})
to compute the evolution of the mode amplitude $c(t)$; and (iii) insert
that mode amplitude $c(t)$ into
Eqs.\ (\ref{nicedotpr2}) and (\ref{nicedotpphi2}) to obtain the
linearized perturbations $\delta p(t)$ and $\delta \phi(t)$ to the
orbital motion.  Steps (i) and (ii) are carried out in Sec.\
\ref{sec:modeevolution}, where we extend previous results in the
vicinity of resonance by Rathore, Blandford and Broderick
\cite{Rathore:2004gs,Rathorethesis}.
Step (iii) is carried out in Sec.\ \ref{sec:phasing}.
We justify the use of this no-backreaction approximation in
Sec.\ \ref{sec:nobcheck}.

\section{Mode amplitude evolution}
\label{sec:modeevolution}

\subsection{Zeroth order solutions}

We start by reviewing the zeroth order solutions of the orbital evolution
equations (\ref{nicedotpr2}) and (\ref{nicedotpphi2}) that apply
in the limit $A_j \rightarrow 0$ of no modal coupling. We denote by $t_0$ the time when mode $j$ enters resonance, and by $\omega_0$
the orbital angular velocity at resonance. We define the radiation reaction timescale at resonance $t_{\rm rr}$
to be
\bea
t_{\rm rr} &=& \frac{5}{256 {\cal M}^{5/3} \omega_0^{8/3}} \nn \\
 &=& 157\, {\rm
  s}
\left(\frac{ {\cal M} }{1.2 \, M_\odot} \right)^{-5/3}
\left(\frac{ f_0 }{10 \, {\rm Hz}} \right)^{-8/3}, \label{trrdef}
\eea
where ${\cal M} = \mu^{3/5} M_{\rm t}^{2/5}$ is the chirp mass and
$f_0 = \omega_0/(2 \pi)$.  We also define the resonance timescale
$t_{\rm res}$ via [cf.\ Sec.\ \ref{sec:estimates} above]
\bea
t_{\rm res} &=& \left( 96 {\cal M}^{5/3}
\omega_0^{11/3}/5\right)^{-1/2} \nn \\
&=& 2.5 \, {\rm s} \left(\frac{ {\cal M} }{1.2 \, M_\odot} \right)^{-5/6}
\left(\frac{ f_0 }{10 \, {\rm Hz}} \right)^{-11/6}.
\label{tresdef}
\eea
We define the dimensionless time parameters
\be
\lambda = \frac{t-t_0}{t_{\rm res}}
\label{eq:lambdadef}
\ee
(time from resonance in units of the resonance time), and
\be
\tau = \frac{t-t_0}{t_{\rm rr}}
\label{eq:taudef}
\ee
(time from resonance in units of the radiation reaction time).  These variables will
be useful in our computations below.
We also define the dimensionless small parameter
\bea
\epsilon &=& \frac{t_{\rm res}}{t_{\rm rr}}
= \frac{32}{3} \sqrt{ \frac{6}{5}} {\cal M}^{5/6} \omega_0^{5/6}
\label{epsilondef} \\
&=& 0.015
\left(\frac{ {\cal M} }{1.2 \, M_\odot} \right)^{5/6}
\left(\frac{ f_0 }{10 \, {\rm Hz}} \right)^{5/6}.
\eea
We will use $\epsilon$ as an expansion parameter throughout our
computations below.
It follows from these definitions that $\epsilon$ is also the ratio of
the orbital and resonance times, up to a constant factor
\be
\omega_0 t_{\rm res} = \frac{8}{3\epsilon},
\label{eq:id1}
\ee
We also have
\be
\tau = \epsilon \lambda.
\label{eq:relation}
\ee

Using these notations, the zeroth order solutions can be written as
\bes\label{oftau}
\bea\label{phioftau}
\phi(t) &=& \phi_0 + \frac{64}{15 \epsilon^2}\Big[1 - (1 -
  \tau)^{5/8}\Big], \\\label{omegaoftau}
\omega(t) &=& {\dot \phi} = \omega_0 (1 - \tau)^{-3/8}, \\
\label{poftau}
p(t) &=& p_0 (1 - \tau)^{1/4}.
\eea
\ees
Here $\phi_0$, $\omega_0$ and $p_0$ are the values of $\phi$, $\omega$
and $p$ at resonance, related by
\be
\omega_0 = \frac{ \sqrt{M_{\rm t}}}{p_0^{3/2}}.
\label{eq:omega0formula}
\ee

\subsection{Mode amplitude evolution}
\label{sec:modeev}

In the rest of this section we will solve
Eq.\ (\ref{nicedotx2}) for the mode amplitude
analytically using matched asymptotic expansions.  We will divide the
inspiral into three regimes: an ``early time'' regime before
resonance, the resonance regime, and a ``late time'' regime after
resonance.  We will compute separate solutions in these three regimes
and match them in their common domains of validity.

The evolution equation (\ref{nicedotx2}) for the mode amplitude can be
written
as
\be
\dot{c}_j + i\omega_j c_j =
A_j(p) \exp\left[im_j\Omega t - im' \phi(t) + i\chi_j \right] ,
\label{nicedotx3}
\ee
where $p(t)$ and $\phi(t)$ are given by Eqs.\ (\ref{oftau}).
It can be seen from this equation that resonance occurs at an orbital
angular frequency $\omega_0$ given by
\be
m' \omega_0 = m_j \Omega + \omega_j.
\label{eq:resonancecondt}
\ee
Note that the right hand side of this equation is just the
inertial-frame frequency of the mode.
We will restrict attention to cases for which the sign of $m'$
is the same as the sign of $m_j \Omega + \omega_j$, so that the
solution $\omega_0$ of Eq.\ (\ref{eq:resonancecondt}) is positive and
resonance does occur.

\subsubsection{Early time solution}
\label{sec:earlytimesoln}

Well before the resonance the amplitude and frequency
of the forcing term on the right hand
side of Eq.\ (\ref{nicedotx3}) are changing on the radiation reaction timescale $t_{\rm
rr}$, which is much larger than the mode period $\sim \omega_j^{-1}$
and the period $\sim \omega^{-1}$ of the tidal forcing.
Therefore to a first approximation we can neglect the time-dependence
of $A_j(t) \equiv A_j[p(t)]$ and $\omega(t) = \dot{\phi}(t)$.  This
gives the following approximate
solution to Eq.\ (\ref{nicedotx3}):
\bea
c_j^e(t) &=& \frac{A_j(t) e^{\left[im_j\Omega t - i m' \phi(t) +
      i\chi_j \right]} }{i[m_j\Omega - m' \omega(t) + \omega_j]}
\nn \\ && \times
\left\{ 1 + O \left[ \frac{ \epsilon^2 \omega_j^2}{ (\omega - \omega_0)^2} \right] \right\},\label{xearly1}
\eea
Here the superscript $e$ denotes "early time", and we have used the
initial condition that the mode excitation vanishes as $t \to
-\infty$. The
error estimate in Eq.\ (\ref{xearly1}) can be obtained
by writing the mode
amplitude as $c^e(t) + \delta c^e(t)$ with $c^e(t)$ given by Eq.\
(\ref{xearly1}), substituting into Eq.\ (\ref{nicedotx3}) and solving for
the linearized correction $\delta c^e(t)$ using the same method of
neglecting the time dependence of $\omega(t)$ and $A(t)$.
The error terms show
that Eq.\ (\ref{xearly1}) is valid in the regime
$| \omega - \omega_0| \gg \epsilon |\omega_j|$.  From
the explicit solution (\ref{omegaoftau}) for $\omega(t)$ this
condition can also be written as
$|\tau | \gg \epsilon$, or, using Eq.\ (\ref{eq:relation}), as $|\lambda|
\gg 1$.  (Recall that $\lambda$ and $\tau$ are negative before resonance).

For later use it will be convenient to specialize the expression
(\ref{xearly1}) for the early time solution to the regime $|t-t_0| \ll
t_{\rm rr}$ or equivalently $|\lambda| \ll 1/\epsilon$.  Expanding the
zeroth order expressions (\ref{oftau}) about $t=t_0$, writing the result in
terms of the rescaled time $\lambda$ and using Eq.\ (\ref{eq:id1})
gives
\bes
\label{eq:expansions}
\bea
\label{eq:mphi}
\phi(t) &=& \phi_0 + \frac{8\lambda}{3\epsilon} + \frac{1}{2}
\lambda^2 + O(\epsilon \lambda^3),\\
\omega(t) &=& \omega_0 \left[ 1 + 3\epsilon \lambda / 8 +
O(\epsilon^2 \lambda^2) \right], \\
A_j(t) &=& A_{0,j} [1 + O(\epsilon \lambda) ]. \label{eq:Atapprox}
\eea
\ees
where $A_{0,j}$ is the value of $A_j$ at resonance. Substituting these
expansions into the early time solution
(\ref{xearly1})
and using Eq.\ (\ref{eq:id1}) and the resonance condition (\ref{eq:resonancecondt})
gives
\bea
c_j^e(\lambda) &=& \frac{8iA_{0,j}}{3 m' \omega_0 \, \epsilon \lambda}
\exp[i(u_r - \omega_j t_{\rm res} \lambda - m' \lambda^2/2)]
\nn \\ && \times
\left[ 1 + O\left(\frac{1}{\lambda^2}\right) + O(\epsilon
\lambda^3) \right],
\label{xearly3}
\eea
where $u_r = \chi_j - m' \phi_0 + m_j\Omega t_0$. Here the error term $O(1/\lambda^2)$ comes from the error term in Eq.\
(\ref{xearly1}), while the second term $O(\epsilon \lambda^3)$ arises from
expanding the exponential in the expression (\ref{eq:mphi}).\footnote{
Since $\lambda \gg 1$ here the $O(\epsilon
\lambda)$ error term from the expansion (\protect{\ref{eq:Atapprox}}) of
the amplitude is negligible in comparison to the $O(\epsilon
\lambda^3)$ error term from the expansion (\protect{\ref{eq:mphi}}) of
the phase.}  The expression (\ref{xearly3}) is therefore valid for $1 \ll
|\lambda| \ll \epsilon^{-1/3}$.

\subsubsection{Late time solution}

The late-time solution after resonance can be found using the same
method as for the early time solution.  The only difference is that we
need to add an arbitrary solution of the homogeneous version of
Eq.\ (\ref{nicedotx3}), which will be determined by matching onto the
solutions in the resonance and early time regions.  The late-time solution can
be written as
\bea
\label{xlate1}
c_j^l(t) &=& \frac{A_j(t) e^{\left[im_j\Omega t - i m' \phi(t) +
      i\chi_j \right]} }{i[m_j\Omega - m' \omega(t) + \omega_j]}
\nn \\ && \times
\left\{ 1 + O \left[ \frac{ \epsilon^2 \omega_j^2}{ (\omega - \omega_0)^2} \right] \right\} + B_l e^{-i[\omega_j(t-t_0) + \chi_l]}, \nn \\
\eea
where $B_l$ and $\chi_l$ are constants parameterizing the homogeneous
solution and
the superscript ``l'' denotes ``late time''.
As before, this solution is valid in the regime $|t-t_0| \gg t_{\rm res}$
or $|\lambda| \gg 1$; in the late time regime $\lambda$ is
positive.  Also as above we can further expand the solution in the
      near-resonance regime $1 \ll |\lambda| \ll \epsilon^{-1/3}$ as
\bea
c_j^l(\lambda) &=& \frac{8iA_{0,j}}{3 m' \omega_0 \, \epsilon \lambda}
\exp[i(u_r - \omega_j t_{\rm res} \lambda - m' \lambda^2/2)]
\nn \\ && \times
\left[ 1 + O\left(\frac{1}{\lambda^2}\right) + O(\epsilon
\lambda^3) \right] + B_l e^{-i(\omega_jt_{\rm{res}} \lambda + \chi_l)}. \nn \\
\label{xlate3}
\eea

\subsubsection{Resonance time solution}

Finally we turn to the resonance regime $|t-t_0| \sim t_{\rm res}$
or $|\lambda| \sim 1$.  In this regime we can expand the differential
equation (\ref{nicedotx3}) using the expansions (\ref{eq:expansions}) and
the formulae (\ref{eq:lambdadef}) and (\ref{eq:id1}) to obtain
\bea
\label{eq:eqres}
\frac{dc^r_j}{d\lambda} + i\omega_j t_{\rm res} c^r_j &=& A_{0,j} t_{\rm res} \times \left[ 1 + O(\epsilon \lambda) \right]
   \\ &&
  \times e^{i\left[ u_r - \omega_j t_{\rm res} \lambda - m' \lambda^2/2 + O(\epsilon \lambda^3) \right]}.\nn
\eea
The solution to this equation consists of a homogeneous solution which
will be determined by matching to the early-time solution, together
with a particular solution which can be expressed in terms of the
Fresnel integrals

\be\label{eq:fresnelapprox}
\left\{\begin{array}{c} C(z) \\ S(z) \end{array}\right\}  = \int_0^z \left\{\begin{array}{c} \cos \\ \sin \end{array}\right\} \left(\frac{\pi u^2}{2}\right)du,
\ee
as discussed by Rathore, Blandford and Broderick
\cite{Rathore:2004gs,Rathorethesis}.
The solution is
\begin{widetext}
\bea\label{xresfinal}
c_j^r(\lambda) &=& \sqrt{\frac{\pi}{|m'|}}A_{0,j} t_{\rm res}
\left[C\left(\sqrt{\frac{|m'|}{\pi}}\lambda\right) - i\,
  \text{sgn}(m')
  S\left(\sqrt{\frac{|m'|}{\pi}}\lambda\right)\right]
e^{i(u_r - \omega_jt_{\rm res} \lambda)}\times[1 + O(\epsilon
  \lambda^2)] + B_re^{-i(\omega_jt_{\rm res}\lambda + \chi_r)}. \nn \\
\eea
\end{widetext}
Here the superscript ``r'' denotes ``resonance'' and $B_r$ and
$\chi_r$ are constants parameterizing the homogeneous part of the solution.
The error term $O(\epsilon \lambda^2)$ in Eq.\
(\ref{xresfinal}) comes from the correction term
in the argument of the exponential in Eq.\ (\ref{eq:eqres}). Note also that
the error terms generated from the errors in the amplitude expansion
(\ref{eq:Atapprox}) scale
as $\epsilon \exp(i\lambda^2)$ or $\epsilon^2 \lambda$ and can be
neglected.
It follows from Eq.\ (\ref{xresfinal})
that this resonance solution is valid in the regime $| \lambda | \ll
\epsilon^{-1/2}$.

For later use it will be convenient to specialize
the expression
(\ref{xresfinal}) for the resonance time solution to the regime
$|\lambda| \gg 1$.  In this regime the arguments of the Fresnel
integrals are large, and we can use the asymptotic formulae
\be
\label{eq:fresnelasymptotic}
\left\{ \begin{array}{c} C(z) \\ S(z) \end{array} \right\} = \frac{{\rm sgn}(z)}{2} + \frac{1}{\pi z} \left\{ \begin{array}{c} \sin \\ -\cos \end{array} \right\} (\pi z^2/2) +
O(z^{-3}).
\ee
This gives
\bea\nn
c_j^r(\lambda) &=& A_{0,j} t_{\rm res}\left[
  \sqrt{\frac{\pi}{2|m'|}} \text{sgn}(\lambda)e^{i(u_r -
    \omega_j t_{\rm res} \lambda - \text{sgn}(m')\pi / 2)}
  \right.
 \nn \\
&& \left. + \frac{i}{m' \lambda}e^{i(u_r - \omega_jt_{\rm res} \lambda
   - m' \lambda^2 / 2)}\right] \times \nn \\
&& [1 + O(\lambda^{-3}) + O(\epsilon \lambda^2)] + B_re^{-i(\omega_jt_{\rm res} \lambda + \chi_r)}. \nn \\
\label{fullxres1}
\eea
This form of the resonance solution is valid in the regime $1 \ll |\lambda| \ll \epsilon^{-1/2}$.

\subsubsection{Matching computation}

We now match the expressions (\ref{xearly3}) for the early time solution and
(\ref{fullxres1}) for the resonance solution within their common domain of
validity $1 \ll |\lambda| \ll \epsilon^{-1/3}$ with $\lambda < 0$.
For both expressions the dominant error terms in $c_j(\lambda)$, which are
obtained by multiplying the fractional errors by the corresponding
expressions for $c_j(\lambda)$, are of order
$O(\lambda^2)$ and $O(\epsilon^{-1} \lambda^{-3})$.
Using Eq.\ (\ref{eq:id1}) one can show that expression (\ref{xearly3})
matches the second term inside the square brackets in Eq.\
(\ref{fullxres1}). Demanding that the
remaining terms in Eq.\ (\ref{fullxres1}) cancel one another determines the constants
$B_r$ and $\chi_r$:
\be
B_r e^{-i \chi_r} = \sqrt{\frac{\pi}{2|m'|}} A_{0,j} t_{\rm res} e^{i (u_r - \text{sgn}(m')\pi/2)} \left[ 1 + O(\epsilon) \right].
\label{eq:Brans}
\ee
Here we have used the fact that ${\rm sgn}(\lambda)= -1$
before resonance. In a similar way we match
the expressions (\ref{xlate3}) for the late time solution and
(\ref{fullxres1}) for the resonance solution within their common domain of
validity $1 \ll |\lambda| \ll \epsilon^{-1/3}$ with $\lambda > 0$.
This yields $B_l = 2 B_r$ and $\chi_l = \chi_r$, or
\be
B_l e^{i \chi_l} = \sqrt{\frac{2\pi}{|m'|}}A_{0,j} t_{\rm res} e^{i (u_r - \text{sgn}(m')\pi/2)} \left[ 1 + O(\epsilon) \right].
\label{eq:Blans}
\ee
The size of $B_l$ implies that at late times when $|\omega -
\omega_j| \gtrsim \omega_j$, the homogeneous term in the
late time solution (\ref{xlate1}) is larger than the particular solution by
a factor $\sim 1/\epsilon$.  Therefore in this regime the late-time
solution is a freely oscillating mode.

\section{Perturbation of orbital motion}
\label{sec:phasing}

In this section we solve the orbital evolution equations
(\ref{nicedotpr2} - \ref{nicedotpphi2}),
treating the tidal terms as linear perturbations on top
of the quasi-circular binary inspiral motion.
We look for solutions of the form
\bes\label{solansatz}
\bea
p(t) &=& p^{(0)}(t) + \delta p(t), \\\label{phiansatz}
\phi(t) &=& \phi^{(0)}(t) +
\delta \phi(t),
\eea
\ees
where $p^{(0)}(t)$ and $\phi^{(0)}(t)$ are the zeroth order
inspiral solutions (\ref{poftau}) and (\ref{phioftau}).
It will be sufficient to specialize to the regime $|\tau| \ll 1$ near resonance.
Linearizing Eqs.\ (\ref{nicedotpr2} - \ref{nicedotpphi2}) yields the
following evolution equations for the perturbations $\delta p$ and
$\delta \phi$:
\bes\label{pertII}
\bea
\label{pertpphiII}
\delta\dot{p} - \frac{3}{4t_\text{rr} (1 - \tau)} \delta p
&=& s(t),\\
\label{pertrII}
{\delta {\dot \phi}} + \frac{3 \sqrt{M_{\rm t}}}{2 p_0^{5/2} (1 -
  \tau)^{5/8}} \delta p&=&0.
\eea
\ees
Here the source term $s(t)$ is
\be
\label{s(t)}
s(t) = 2 s_0 A_{0,j} \Re \left[ c_j^* e^{- i m' \phi(t) + i m_j \Omega t + i \chi_j}
  \right] \, \left[ 1 + O(\epsilon \lambda) \right],
\ee
where
\be
s_0 = - \frac{2 m' b_j \sqrt{p_0}}{\mu \sqrt{M_{\rm t}}}.
\label{s_0}
\ee

\subsection{Evolution of semi-latus rectum}
\label{sec:evol1}

Equation (\ref{pertpphiII}) is a first order differential equation and
its solution is easily found to be
\be
\delta p = \int_{-\infty}^t s(t^\prime) \left(\frac{1 -
  \tau^\prime}{1 - \tau}\right)^{3/4} \, dt^\prime,
\label{eq:deltapformula}
\ee
where $\tau^\prime = t^\prime / t_{\rm rr}$.
In evaluating integral (\ref{eq:deltapformula}), we use the early
time approximation (\ref{xearly1}) for the mode amplitude $c(t)$ for
$-\infty < t < t_0-t_{\rm m}$, the resonance time approximation
(\ref{xresfinal}) for $t_0-t_{\rm m} < t < t_0+t_{\rm m}$, and the late time
approximation (\ref{xlate1}) for $t > t_0+t_{\rm m}$.  We denote by
$s^e(t)$, $s^r(t)$, and $s^l(t)$ the corresponding approximations to
the driving term (\ref{s(t)}), and by $\delta p^e(t)$, $\delta p^r(t)$ and
$\delta p^l(t)$ the corresponding approximate solutions for $\delta p$
in the three regimes.  We choose the parameter $t_m$ governing the times
at which we switch
from one approximation to the next to be
\be
t_{\rm m} = \sigma t_{\rm rr} \epsilon^{4/5},
\label{eq:tmatch}
\ee
where $\sigma$ is a dimensionless constant of order unity.
The corresponding values of $\tau = (t-t_0)/t_{\rm rr}$ and $\lambda
=\tau/\epsilon$ are $\tau_{\rm m} = \sigma \epsilon^{4/5}$ and $\lambda_{\rm
m} = \sigma \epsilon^{-1/5}$.  The choice (\ref{eq:tmatch}) of scaling
with $\epsilon$ of $t_{\rm m}$ maximizes the
overall accuracy, as can be seen from the scalings of the error
estimates in the mode amplitude solutions (\ref{xearly3}), (\ref{xlate3}) and
(\ref{fullxres1}).  The final results are independent of the choice of
$\sigma$.

Performing integral (\ref{eq:deltapformula}) is straightforward,
albeit a little tedious. The result of interest to us here is the
expansion of $\delta p^l(t)$ in the regime $1 \ll \lambda  \ll
\epsilon^{-1/2}$. This is
\begin{widetext}
\bea
\delta p^l(t)&=& \frac{2\pi}{|m'|} s_0 A_{0,j}^2
t_{\text{res}}^2\left[1 - \frac{1}{\lambda}\sqrt{\frac{2}{\pi
      |m'|}}\cos\left(\frac{|m'|\lambda^2}{2} +
  \frac{\pi}{4}\right) + \frac{1}{2\pi |m'|\lambda^2} +
  O(\lambda^{-3}) + O(\epsilon \lambda^2) \right].
\label{eq:deltaplate3}
\eea
\end{widetext}

\subsection{Phase evolution}
\label{sec:evol2}

We obtain the phase perturbation $\delta \phi$ by inserting into
Eq.\ (\ref{pertrII}) the perturbation
$\delta p$ of the semi-latus rectum and integrating.
We first discuss the leading order phase shift and then discuss the
magnitude of the corrections.

To leading order we take $\delta p(t) = 0$ for $t<t_0$, and for $t>t_0$ we
use the expression for $\delta p(t)$ given by the first term in the brackets in Eq.\ (\ref{eq:deltaplate3}). This approximate form of $\delta
p(t)$ satisfies for $t>t_0$ the homogeneous version of the equations of motion
(\ref{pertII}).  It coincides at $|\tau| \ll 1$ with the homogeneous solution
obtained by taking $t_0 \to t_0 - \Delta t$ in the zeroth-order
solution given by Eqs.\ (\ref{eq:taudef}) and (\ref{poftau}) and
linearizing in $\Delta t$:
\be\label{eq:plinearized}
\delta p(t) = - \frac{ p_0 \Delta t} {4 (1-\tau)^{3/4} t_{\rm rr}}.
\ee
By expanding Eq.\ (\ref{eq:plinearized}) in the regime $1 \ll \lambda
\ll \epsilon^{-1/2}$, and comparing with the first term in Eq.\
(\ref{eq:deltaplate3}), we
can read off the parameter $\Delta t$:
\be
\Delta t =  -\frac{8\pi}{|m'| p_0}s_0 A_{0,j}^2 t_{\text{res}}^2t_{\rm rr}\left[ 1 + O(\epsilon) \right].
\label{eq:Deltatans}
\ee
We discuss below the $O(\epsilon)$ error term.

The corresponding solution for the phase perturbation $\delta \phi$ is also a
solution of the homogeneous system of equations, and therefore must be
of the form discussed in the introduction, parameterized by the time
delay parameter $\Delta t$ and an overall phase shift $\Delta \phi$:
\be\label{eq:phiapproxprime}
\phi(t) = \phi^{(0)}(t) + \delta \phi(t)  = \phi^{(0)}(t + \Delta t) -
\Delta \phi.
\ee
In this approximation the relation
(\ref{eq:constraint}) between $\Delta \phi$ and
$\Delta t$ follows from the continuity
of $\delta \phi(t)$ at $t=t_0$, which follows from Eq.\ (\ref{pertrII}):
\be
\Delta \phi = \omega_0 \Delta t.
\label{eq:qq}
\ee
Combining this with the expression (\ref{eq:Deltatans}) for $\Delta
t$ and the formula (\ref{s_0}) for $s_0$
gives the phase shift formula\footnote{The gravitational wave
 phase shift $\Delta \Phi$ used in the introduction is related to the
 orbital phase shift $\Delta \phi$ used here by $\Delta \Phi = 2
 \Delta \phi$.}
\be
\Delta \phi = 16\pi \,
 \text{sgn}(m') \, \frac{ A_{0,j}^2 b_j
t_{\text{res}}^2 t_{\rm rr}}{\mu p_0^2}  \left[ 1 + O(\epsilon) \right].
\label{answers}
\ee

We next discuss the sign of the phase shift $\Delta \phi$.  We shall
see later [cf.\ Eqs.\ (\ref{eq:qq}) and (\ref{eq:ec1})] that the sign of the phase
shift is the same of the sign of the net energy transferred from the
orbit to the star.  The sign of the parameter $b_j$ in Eq.\ (\ref{answers})
coincides with the sign of $\omega_j$ for inertial modes, and probably
also in general \cite{paperI}.  Therefore we obtain
\bea
\text{sgn}(\Delta \phi) &=& \text{sgn}(m' \omega_j) = \text{sgn}(m_j
\Omega + \omega_j) \, \text{sgn}(\omega_j) \nn \\
&=& \text{sgn}(\Omega +\omega_j/m_j) \, \text{sgn}(\omega_j/m_j),
\eea
where on the first line we have used the resonance condition
(\ref{eq:resonancecondt}).
Since $\Omega>0$ by convention it follows that $\Delta \phi$ is
negative if and only if both $\omega_j/m_j$
is negative (the mode is retrograde in the corotating frame) and
$\omega_j/m_j + \Omega$ is positive (the mode is prograde in the
inertial frame), i.e., the Chandrasekhar-Friedman-Schutz mode
instability criterion \cite{1970PhRvL..24..762C,1978ApJ...222..281F}
is satisfied.

Consider now the magnitude of the corrections to the leading order
result (\ref{answers}).  First, there is a nonzero phase perturbation in the early
time regime, generated by the perturbation $\delta p$.
This phase perturbation grows like $1/|\lambda|$, but is smaller than
the phase shift (\ref{answers}) by a factor of $\sim \epsilon$ when
the growth saturates at $|\lambda| \sim 1$.  Second, there is a phase
perturbation which is
generated by inserting the second and third terms in the
approximation (\ref{eq:deltaplate3})
for $\delta p^r(t)$, and integrating using
Eq.\ (\ref{pertrII}).  This phase perturbation is also is smaller than
$\Delta \phi$ by a factor of $\sim \epsilon$.
Third, there are fractional corrections of order $O(\epsilon)$ due to
higher order contributions to the matching parameters $B_r$, $\chi_r$,
$B_l$ and $\chi_l$ indicated by the error terms in Eqs.\ (\ref{eq:Brans}) and
(\ref{eq:Blans}).
Finally, there are phase perturbations generated at late times by the coupling between the freely oscillating mode and orbital motion. These can also be shown to be smaller than the leading order $\Delta \phi$ by a factor of at most $\sim \epsilon$.

\subsection{Validity of circular approximation}
\label{sec:circularcheck}

When solving the orbital evolution equations (\ref{eq:pdot00}) and
(\ref{eq:phidot00}) we set the eccentricity $e$ to zero.  We now
determine the domain of validity of this approximation.

First, in the absence of mode coupling, there is a non-zero
instantaneous eccentricity $e \sim t_{\rm orb} / t_{\rm rr} \sim
\epsilon^2$ due to the gradual inspiral.  To compute this eccentricity
we start with the orbital equations of motion (\ref{dotpr}) and (\ref{dotpphi})
with the mode coupling terms dropped, and solve using a two timescale
expansion.  The result is that the zeroth order inspiral solution
(\ref{oftau}) is accurate up to fractional
corrections of order $O(\epsilon^4)$:
\bes
\bea
r(t) &=& r_0 (1 -\tau)^{1/4} \left[1 + O(\epsilon^4) \right], \\
\phi(t) &=& \int dt \, \omega_0 (1 - \tau)^{-3/8} \left[1 + O(\epsilon^4) \right].
\eea
\ees
Transforming to the variables $p(t)$, $e(t)$, $\phi(t)$ and
$\phi_0(t)$ using the definition (\ref{eq:newvars}) gives
\bes
\bea
p(t) &=& r_0 (1 -\tau)^{1/4} \left[1 + O(\epsilon^4) \right], \\
e(t) &=& \frac{1}{4 t_{\rm rr} \omega_0} (1 - \tau)^{-5/8} \left[1 +
  O(\epsilon^2) \right], \\
\phi(t) &=& \int dt \, \omega_0 (1 - \tau)^{-3/8} \left[1 +
  O(\epsilon^4) \right], \\
\phi_0(t) &=& \left[ \phi(t) - 3 \pi/2 \right] \left[ 1 + O(\epsilon^2) \right].
\eea
\ees

We now denote this solution by $p^{(0)}(t)$, $e^{(0)}(t)$,
$\phi^{(0)}(t)$ and $\phi_0^{(0)}(t)$, and linearize the equations
(\ref{eq:newvars1}) to solve for the perturbations $\delta p(t)$,
$\delta e(t)$, $\delta \phi(t)$ and $\delta \phi_0(t)$ generated by
the mode coupling.  We retain only terms that are zeroth order in $\epsilon$.
The resulting equation for $\delta e(t)$ is
\be
\dot{\delta e} = - \frac{ \sqrt{p^{(0)}} }{ M_{\rm t} } a^{\rm mode}_{\hat r},
\label{eq:dotdeltae}
\ee
where $a^{\rm mode}_{\hat r}$ is the radial acceleration due to the
mode coupling, given by Eqs.\ (\ref{eq:arhat}) and (\ref{eq:fjformula}).
We now use the expression (\ref{xresfinal}) for
the mode amplitude in the resonant regime, simplify using the
expansions (\ref{eq:expansions}), and in the resulting expression drop
all the terms that do not accumulate secularly.  The result is
\be
\delta e(t) = \delta e_{\rm max} {\cal J}(\sqrt{|m'|} \lambda),
\ee
where the function ${\cal J}(\lambda)$ is given by
\bea
{\cal J}(\lambda) &\equiv & \int_{-\infty}^\lambda d\bar{\lambda} \
\bigg\{ -\sin(\bar{\lambda}^2/2) \left[
C\left(\frac{\bar{\lambda}}{\sqrt{\pi}}\right) + \frac{1}{2} \right]
\nn \\
&&+ \cos(\bar{\lambda}^2/2) \left[
S\left(\frac{\bar{\lambda}}{\sqrt{\pi}}\right) + \frac{1}{2} \right] \bigg\}.
\nn
\eea
This function satisfies ${\cal J}(\lambda) \to 0$ as $\lambda \to
\infty$, from the fact that the Fresnel functions $C$ and $S$ are odd
and using the approximate formulae (\ref{eq:fresnelasymptotic}).
Therefore in this approximation the final eccentricity generated by
the resonance vanishes, in agreement with a previous result of Rathore
[Eq.\ (6.104) of Ref.\ \cite{Rathorethesis} specialized to $e=0$].

The quantity $\delta e_{\rm max}$ is, up to a factor of order unity,
the maximum eccentricity achieved during the resonance.  It scales as
\be
\delta e_{\rm max} \sim
\epsilon^2 \Delta \phi.
\ee
This eccentricity is transient and lasts
only for a time $\sim t_{\rm res}$.
From the equations of motion (\ref{eq:pdot00}) and (\ref{eq:phidot00})
we can estimate the corrections to the parameters
$\Delta \phi$ and $\Delta t$ characterizing the resonance that are
generated by this transient eccentricity.  We find that the corrections to both
$\Delta \phi$ and $\omega_0 \Delta t$ are of order $\epsilon \Delta
\phi$ [$\delta e_{\rm max}$ times the number of cycles $\sim
1/\epsilon$ of resonance] or smaller, and thus can be neglected.

We note that this computation is valid only in the regime $\Delta \phi
\ll 1$ for which
$\delta e \ll e^{(0)}$ throughout the resonance.  In the regime
$\Delta \phi \gg 1$, both $e(t)$ and $\phi_0(t)$ are
significantly perturbed away from their zeroth order solutions, and it
is not a good approximation to evaluate the right hand side of Eq.\
(\ref{eq:dotdeltae}) using the zeroth order solutions as done here.
We conclude that the circular approximation is valid for this paper, since
$\Delta \phi \alt 1$ for all the cases we consider.

\subsection{Validity of no-backreaction approximation}
\label{sec:nobcheck}

The no-backreaction approximation is valid when the phase perturbation
$\delta \phi(t)$ accumulated over the resonance is small compared to
the phase accumulated over the same period in the absence of mode
coupling \cite{Rathorethesis}, i.e., when
\be\label{eq:crit}
\Upsilon \equiv \frac{\delta\phi(t_0 + t_{\rm res}) - \delta
\phi(t_0-t_{\rm res})}{\omega_0 t_{\rm res}} \ll 1.
\ee
This ensures the smallness of the backreaction fractional corrections
to the mode driving terms on the right hand side of Eq.\ (\ref{nicedotx3}).
From Eq.\ (\ref{pertrII})
we have
\be
\delta\phi(t_0 + t_{\rm res}) - \delta\phi(t_0 - t_{\rm res})
\sim \frac{\sqrt{M_{\rm t}}}{p_0^{5/2}}
t_{\rm res} \delta p(t_0).
\ee
Using Eq.\ (\ref{eq:deltapformula})
to estimate $\delta p(t_0)$ now gives
\be
\Upsilon \sim \frac{t_{\rm rr} A_{0,j}^2 b_j}{\mu p_0^2 \omega_0^2}.
\ee
Comparing this with the formula (\ref{answers}) for the resonance
phase shift $\Delta \phi$ and using Eq.\ (\ref{eq:id1}) gives
\be
\Upsilon \sim \epsilon^2 \Delta \phi.
\label{eq:Upsilonestimate}
\ee

The estimate (\ref{eq:Upsilonestimate}) shows that in general there is
a non-empty
regime where the no-backreaction approximation is valid, $\Upsilon \ll
1$, and where in addition the phase shift due to the resonance is
large, $\Delta \phi \gg 1$.  For the {\it r}-modes studied here, by
combining the numerical estimate (\ref{eq:ans2ss}) of $\Delta \phi$ with the
numerical estimate (\ref{epsilondef}) of $\epsilon$ we obtain
\be
\Upsilon \sim 10^{-3} \, M_{1.4}^{-5/3} f_{s100}^{7/3} R_{10}^4.
\ee
Thus the no-backreaction approximation is valid for the mode driving
considered here.

\section{Application to gravitomagnetic tidal driving of Rossby modes}
\label{sec:rmodedriving}

\subsection{Overview}
\label{sec:rmodeoverview}

In this section we compute the parameters $\Delta t$ and $\Delta \Phi$
that characterize the perturbation (\ref{eq:signature})
to the gravitational wave phase due to the resonant post-1-Newtonian tidal driving
of \textit{r}-modes.

In our computations we will use a harmonic, conformally Cartesian
post-1-Newtonian coordinate system adapted to the star whose modes are
being driven.  We can specialize this coordinate system so that
(i) the post-1-Newtonian mass dipole of the star
vanishes, so that the origin of coordinates coincides with the star's
center of mass; (ii) the angular velocity of the
coordinate system as measured locally using Coriolis-type accelerations
vanishes (the angular velocity with respect to distant stars will not
vanish due to dragging of inertial frames); and (iii)
the normalization of the time coordinate is such that the $l=0$ piece
of the Newtonian potential outside the star is of the form $\Phi = -M/r$
without any additional additive constant \cite{1991PhRvD..43.3273D,paperI}.
In such body-adapted reference frames (also called local asymptotic
rest frames)
the effect of the external gravitational field on the internal
dynamics of the star can be parameterized, in post-1-Newtonian
gravity, by a set of gravitoelectric tidal tensors
$\mathcal{E}_{i_1 i_2 \ldots i_l}(t)$ and gravitomagnetic tidal tensors
$\mathcal{B}_{i_1 i_2 \ldots i_l}(t)$, for $l \ge 2$.
These tensors are symmetric and tracefree on all pairs of indices, and
are invariant under the remaining post-1-Newtonian gauge freedom \cite{paperI}.

In Newtonian gravity the gravitoelectric tidal fields are just the
gradients of the external potential $\Phi_{\rm ext}$ evaluated at the
spatial origin:
\be
{\cal E}_{i_1 \ldots i_l}(t) = \frac{1}{(l-2)!} \frac{ \partial^l \Phi_{\rm
    ext}}{\partial x^{i_1} \ldots \partial x^{i_l} }(\bm{0},t).
\ee
In post-1-Newtonian gravity the definition of the tidal tensors is more
complicated, since the field equations are nonlinear and so the
potentials can not be written as the sum of interior and exterior
pieces as in the Newtonian case.  The post-1-Newtonian definitions are
discussed in detail in Refs.\
\cite{1991PhRvD..43.3273D,paperI}.\footnote{
We follow here the notational conventions of Ref.\ \cite{Thorne:1984mz}
for the tidal tensors
${\cal E}_{i_1\ldots i_l}(t)$ and ${\cal B}_{i_1 \ldots i_l}(t)$.
A different notational convention is used in Refs.\
\protect{\cite{1991PhRvD..43.3273D,paperI}}, where these tensors are denoted
$- G_{i_1 \ldots i_l}(t)/ (l-2)!$ and $- 3 H_{i_1 \ldots i_l}(t) / [2
(l+1) (l-2)!]$ respectively.}
The tidal tensors ${\cal E}_{i_1 i_2 \ldots i_l}$ and ${\cal B}_{i_1
i_2 \ldots i_l}$ can also be defined in the more general context of a
small object in an arbitrary background metric \cite{Thorne:1984mz}.

In this section we will compute the driving of the {\it r}-modes by
the leading order gravitomagnetic tidal tensor ${\cal B}_{ij}(t)$.
The effects of the higher order tensors ${\cal B}_{i_1 \ldots i_l}(t)$
for $l \ge 3$ are suppressed by one or more powers of the tidal
expansion parameter $R/r$, where $R$ is the radius of the star and $r$
is the orbital separation.  Therefore we will neglect these higher
order tensors.

A key point about the gravitomagnetic tidal tensors is that they
vanish identically at Newtonian order.  Therefore when computing the
driving of the {\it r}-modes, it is sufficient to use
Newtonian-order stellar perturbation theory supplemented by
post-1-Newtonian tidal driving terms.
This is sufficient to compute the mode driving to post-1-Newtonian
accuracy, and constitutes a significant simplification.\footnote{Note
that the validity of this approximation depends on the choice of
coordinate system.  It is valid for the coordinate systems used here,
but would not be valid for a coordinate system $(t,x^i)$
obtained from the standard global inertial-frame harmonic coordinate system
$({\tilde t},{\tilde x}^i)$ by a Newtonian-type transformation
of the form $t = {\tilde t}$, $x^i = {\tilde x}^i - z^i({\tilde t})$.}

For the Newtonian mode driving discussed in the previous sections, we used
the Hamiltonian (\ref{eq:Hamiltonian}) to deduce driving terms
(\ref{eq:amodesformula}) in the orbital equations of motion.  Here,
for the post-Newtonian case, we will instead compute those driving terms
directly, by evaluating the current quadrupole $S_{ij}$ induced by the r-mode
and using the post-1-Newtonian equations of motion including multipole
couplings derived in Ref.\ \cite{paperI}.  We will verify that these
driving terms take the same form (\ref{eq:ahatphi1}) as in the Newtonian case and
as analyzed in Secs.\ \ref{sec:modeevolution} and \ref{sec:phasing}.

The remainder of this section is organized as follows.
We review the properties of {\it r}-modes
in Sec.\ \ref{sec:rmodes}.  Section \ref{sec:forceterms} and Appendix
\ref{app:Y} compute the
mode-orbit coupling term responsible for driving the mode, as well as the resonant response of the modes.
Next in Sec.\ \ref{sec:backreaction} we
analyze the effect of the mode on the orbit and
derive the parameters $\Delta t$ and $\Delta \Phi$, using the results of
Sec.\ \ref{sec:phasing}.

\subsection{Rossby modes}
\label{sec:rmodes}

In this section we review the relevant properties of Rossby modes ($r$-modes).
We will use the notations of Sec.\ \ref{sec:modeexpand} above, in particular the
inertial frame coordinates are ${\bm x} = (r,\theta_s,\phi_s)$ and the
coordinates that co-rotate with the star are ${\bar {\bm x}} =
(r,\theta_s,{\bar \phi}_s)$ with $\phi_s = {\bar \phi}_s + \Omega
t$.  In the remainder of this section we will drop the mode index $j$ for
simplicity, and we will write the rotating-frame mode frequency
$\omega_{j}$ as $\omega_{\rm m}$.

For {\it r}-modes, the rotating frame
mode frequencies are \cite{1981A&A....94..126P}
\be\label{modefreq}
\omega_{\rm m} = \omega_{lm} = -\frac{2m\Omega}{l(l+1)},
\ee
where $\Omega = |{\bf \Omega}|$ and $l,m$ are the usual
spherical harmonic indices. The mode eigenfunctions are \cite{1981A&A....94..126P}
\be\label{eq:rmodeeigenfunction}
\bm{\xi}_{lm}(\bar{\bm{x}}) = -\frac{i f_{lm}(r)}{\sqrt{l(l+1)}} \, \bar{\bm {x}} \times
\bm{\nabla} Y_{lm}(\theta,\bar{\phi}) \left[ 1 + O(\varpi^2) \right],
\ee
where $Y_{lm}$ is a spherical harmonic and $f_{lm}(r)$ is a real radial mode
function.  Here $\varpi = \Omega/\Omega_{\rm cr}$ is the angular
velocity in units of the break-up angular velocity $\Omega_{\rm cr} =
\sqrt{M/R^3}$.
The mode normalization condition (\ref{modenorm})
can be written in terms of the radial mode function $f_{lm}(r)$ as
\be
\int_0^R dr \, r^2 \rho(r)  f_{lm}(r)^2 = M R^2.
\label{modenorm1}
\ee
Using the mode function (\ref{eq:rmodeeigenfunction}) in the
definition (\ref{eq:bdef}) of the constant $b$ yields
\be
b = \omega_{\rm m} M R^2.
\label{eq:bans}
\ee
We define a dimensionless coupling parameter $I_{lm}$ for the mode
by
\be\label{eq:radialint}
I_{lm} = \frac{1}{M R^3}\int_0^R r^4 f_{lm}(r) \, \rho(r) \, dr;
\ee
this parameter will arise in our computations below.

We next express the mode function
(\ref{eq:rmodeeigenfunction}) in inertial coordinates
$({\bm x},t)$, in which it simply acquires
a factor of $e^{-im\Omega t}$:
\be
\label{eq:rmodeeigenfunction1}
\bm{\xi}_{lm}(\bm{x}) = -\frac{i e^{-im\Omega t}}{\sqrt{l(l+1)}}
f_{lm}\, \bm {x} \times
\bm{\nabla} Y_{lm}(\theta,\phi) \left[ 1 + O(\varpi^2) \right].
\ee
For each $l$, $m$ we define the symmetric tracefree tensors
$\mathcal{Y}^{lm}_{s_1s_2
  ... s_l}$ by the equation \cite{thorne}
\be
Y_{lm}(\theta,\phi) = \mathcal{Y}_{s_1s_2 ... s_l}^{lm}
n^{s_1} n^{s_2} \ldots n^{s_l},
\ee
where $n^i = x^i/r$ is the unit radial vector.

\subsection{Computation of force terms and resonant response of modes}
\label{sec:forceterms}

We now turn to the analysis of the mode amplitude driving terms
$\langle \bm{\xi}_{lm} , \bm{a}_{\text{ext}} \rangle$.
We restrict attention to modes $\bm{\xi}_{lm}$ with $m > 0$; the $m<0$
driving terms are just the complex conjugates of the $m>0$ terms.
In order to derive
the form of the external acceleration $\bm{a}_{\text{ext}}(\bm{x},t)$ to
be added to the Newtonian perturbation equations, we consider,
temporarily, post-1-Newtonian stellar perturbation theory.
The argument which follows is a slightly more rigorous version of the
argument given in Appendix A of Ref.\ \cite{Favata:2005da}.

In post-1-Newtonian gravity in conformally Cartesian gauges, the
metric is expanded in the form
\begin{eqnarray}\nn
ds^2 &=& - [1 + 2\varepsilon^2\Phi + 2\varepsilon^4(\Phi^2 + \psi) +
O(\varepsilon^6)](dt/\varepsilon)^2 \\\nn
& & + [2\varepsilon^3\zeta_i + O(\varepsilon^5)]dx^i(dt/\varepsilon)
\\\label{metric}
& & + [\delta_{ij} -2\varepsilon^2\Phi \delta_{ij} +
O(\varepsilon^4)]dx^idx^j.
\label{metric0}
\end{eqnarray}
Here $\Phi$, $\psi$ and $\bm{\zeta}$
are the Newtonian potential, the post-Newtonian scalar
potential and the gravitomagnetic potential, respectively.  The
quantity $\varepsilon$ is a formal expansion parameter which can be
set to unity at the end of the calculation; it can be thought of as
the reciprocal of the speed of light.

We consider a uniformly rotating background star, characterized by a
pressure $p_0$, density $\rho_0$, fluid 3-velocity $\bm{v}_0$, and by
gravitational potentials $\Phi_0$, $\psi_0$, and $\bm{\zeta}_0$.
We assume that the star is subject to an external perturbing
gravitomagnetic tidal field ${\cal B}_{ij}(t)$.  The linear
response of the star can be parameterized in terms of the Eulerian
perturbations $\delta p$, $\delta \rho$, $\delta \bm{v}$, $\delta
\Phi$, $\delta \psi$, and $\delta \bm{\zeta}$, which satisfy the
linearized post-1-Newtonian hydrodynamic and Einstein equations.
The solution is determined by the boundary conditions on the
gravitational potential perturbations at large $r$, which are
\bes\label{eq:calBbc}
\bea
\label{eq:zetabc}
\delta \zeta_i(\bm{x},t) &\to& \frac{2}{3} \epsilon_{ijk}
       {\cal B}_{kl}(t) x^j x^l \\
\delta \Phi(\bm{x},t) &\to& 0 \\
\delta \psi(\bm{x},t) &\to& 0
\label{eq:psibc}
\eea
\ees
as $r \to \infty$.
[See, for example, Eqs.\ (3.5) of Ref.\ \cite{paperI}, where we drop
all the tidal tensors except for $H_{ij} = -2 {\cal B}_{ij}$.]
Next, the linearized equation satisfied by $\delta \bm{\zeta}$ in
harmonic gauges is
\be
\nabla^2 \, \delta\zeta^i - 16\pi \rho_0 \delta v^i
- 16\pi v_0^i  \delta \rho =0.
\label{eq:deltazetaeqn}
\ee
We can neglect the second and third terms in this equation for the
reason explained in Sec.\ \ref{sec:rmodeoverview} above: the
perturbations $\delta \rho$ and $\delta \bm{v}$ will be proportional
to ${\cal B}_{ij}$ and thus of post-1-Newtonian order, so the
corrections to $\delta \bm{\zeta}$ they generate will be of
post-2-Newtonian order.  Solving Eq.\ (\ref{eq:deltazetaeqn})
without the second and third terms and subject to the boundary condition
(\ref{eq:zetabc}) yields
\be
\delta \zeta_i(\bm{x},t) = \frac{2}{3} \epsilon_{ijk}
       {\cal B}_{kl}(t) x^j x^l.
\label{eq:deltazetaans}
\ee
Also, since the perturbation has no Newtonian part, we have
\be
\delta \Phi=0.
\label{eq:Phizero}
\ee

Next, we take the field equation for $\psi$, the post-1-Newtonian continuity
equation, and the post-1-Newtonian Euler
equation in harmonic gauge.  We linearize these equations about the
background solution
$(\rho_0,p_0,\bm{v}_0,\Phi_0,\psi_0,\bm{\zeta}_0)$, use Eq.\ (\ref{eq:Phizero})
and drop all terms of the type discussed above that give rise to
post-2-Newtonian corrections.  The results are
\be
\nabla^2 \delta \psi = 4 \pi \delta \rho,
\label{eq:psi5}
\ee
\be
\dot{\delta \rho} + \bm{\nabla} \cdot ( \delta \rho \bm{v}_0 + \rho_0
\delta \bm{v} ) =0,
\label{eq:continuity}
\ee
and
\bea
\dot{\delta \bm{v}} &=& -(\bm{v}_0 \cdot \bm{\nabla}) \delta \bm{v}
- (\delta \bm{v} \cdot \bm{\nabla}) \bm{v}_0  - \frac{\bm{\nabla} \delta
p}{\rho_0} + \frac{\bm{\nabla} p_0}{\rho_0^2} \delta \rho \nn \\
&& - \bm{\nabla} \delta \psi + \bm{a}_{\rm ext},
\label{eq:Euler1}
\eea
where
\be\label{aextgm}
\bm{a}_\text{ext} = - \dot{\delta \bm{\zeta}} +
\bm{v}_0\times(\nabla\times\delta\bm{\zeta}).
\ee
Finally, we can replace the post-1-Newtonian background solution
$(\rho_0, p_0, \Phi_0,\bm{v}_0)$ by its Newtonian counterpart; the
corresponding changes to $\delta \rho$, $\delta p$, $\delta \bm{v}$ and $\delta \psi$
are of post-2-Newtonian order and can be neglected.

Equations (\ref{eq:psi5}), (\ref{eq:continuity}) and (\ref{eq:Euler1})
together with the boundary condition (\ref{eq:psibc})
are precisely the standard Newtonian perturbation equations
supplemented by the external acceleration $\bm{a}_{\rm ext}$.
We have therefore shown that the leading order effect of the external
gravitomagnetic tidal field ${\cal B}_{ij}(t)$ on the star can be
computed using Newtonian perturbation theory.
The expression (\ref{aextgm}) for the external acceleration
$\bm{a}_{\rm ext}$ can be
rewritten using the formula (\ref{eq:deltazetaans}) for $\delta \bm{\zeta}$ as
\bea
a_{\rm ext}^i &=& - \frac{2}{3} \epsilon_{ijk} \dot{{\cal
    B}}_{kl}(t) x^j x^l
-2 \epsilon_{ijk} v^0_j {\cal B}_{kl}(t) x^l.
\eea
Using this expression together with the mode function
(\ref{eq:rmodeeigenfunction1}) and $\bm{v}_0 = \bm{\Omega} \times
\bm{x}$ in the coupling integral (\ref{modedrivingterm}) written in inertial frame
coordinates shows that the only nonvanishing driving terms occur for
$l=2$, and we obtain
\bea
\langle \bm{\xi}_{2m} , \bm{a}_{\text{ext}} \rangle &=& - \frac{ 16 \pi
  i M R^3 I_{2m}}{15 \sqrt{6}} e^{i m \Omega t} (
\mathcal{Y}^{2m}_{ij})^* \nn \\
&& \times \left[ {\dot {\cal B}}_{ij} + 2 {\cal B}_{pi} \epsilon_{jpq}
  \Omega_q \right] \nn \\
&=& - \frac{ 16 \pi
  i M R^3 I_{2m}}{15 \sqrt{6}}
\frac{d}{dt} \left[ e^{i m \Omega t} (\mathcal{Y}^{2m}_{ij})^*
  {\cal B}_{ij}(t)\right]. \nn \\
\label{eq:drivingans}
\eea
Here $I_{2m}$ are the dimensionless coupling parameters
  (\ref{eq:radialint}), and we have used $\bm{\Omega} = (0,0,\Omega)$.

So far the analysis has been valid for a star placed in an arbitrary
gravitomagnetic tidal field ${\cal B}_{ij}(t)$.  We now specialize to
a star in a binary.  We denote by $\bm{z}(t)$ and $\dot{\bm{z}}(t)$
the coordinate location and velocity of the companion star of mass
$M^\prime$; these quantities are the relative displacement and
relative velocity since we are working in the center of mass frame of
the star $M$.  In Appendix \ref{app:Y} we show that
\be
{\cal B}_{ij} = 6 M^\prime z_{(i} \epsilon_{j)kl} z_k \dot{z}_l / r^5.
\label{eq:calBans}
\ee
We parameterize the quasi-circular orbit as
\be
\bm{z}(t) = r(t) \, [\cos \psi \cos\phi(t), \sin\phi(t), \sin \psi
  \cos \phi(t)].\,\,\,
\label{eq:orbitp}
\ee
Here $\phi(t)$ is the orbital phase of the binary, and  $\psi$ is
the inclination angle of the orbital angular momentum relative to
the spin axis of the star.  Inserting this parameterization into the
formula (\ref{eq:calBans}) for ${\cal B}_{ij}$ and then into the
expression (\ref{eq:drivingans}) for the overlap integral
gives the following
results for the $l=2,m=1$ and $l=2,m=2$ {\it r}-modes
\begin{widetext}
\bes\label{forcingsII}
\bea
\langle \bm{\xi}_{21}, \bm{a}_\text{ext} \rangle &=&
\sqrt{\frac{16\pi}{5}}\frac{I_{21}}{r^2} M M^\prime R^3 \omega e^{i\Omega
  t}
    \left[(\omega +
\Omega)\sin\left(\frac{3\psi}{2}\right)\sin\left(\frac{\psi}{2}\right)e^{i\phi(t)}
 + (\omega - \Omega)
\cos\left(\frac{3\psi}{2}\right)\cos\left(\frac{\psi}{2}\right)
e^{-i\phi(t)}\right],
\nn \\ && \, \\
\langle \bm{\xi}_{22}, \bm{a}_\text{ext} \rangle &=&
\sqrt{\frac{16\pi}{5}}\frac{I_{22}}{r^2} M M^\prime R^3 \omega
e^{i2\Omega t} \sin\psi
  \left[(2\Omega +
  \omega)\sin^2\left(\frac{\psi}{2}\right)e^{i\phi(t)} - (2\Omega -
  \omega) \cos^2\left(\frac{\psi}{2}\right)e^{-i\phi(t)}\right],
\eea
\ees
\end{widetext}
where $\omega = {\dot \phi}$.

These driving terms are each a sum of two terms proportional to $e^{ i m \Omega
  t \pm i \phi(t)}$  which oscillate
at different frequencies.  From the equation of motion
(\ref{dotxrmode}), resonant driving will occur at $\omega(t) =
\omega_0$ if
\be
- \omega_{\rm m} = m \Omega \pm \omega_0.
\ee
Now the mode frequency $\omega_{\rm m}$ is negative and by Eq.\
(\ref{modefreq}) satisfies $|\omega_{\rm m}| < \Omega$.  Also the
orbital angular velocity $\omega(t)$ is by convention positive.
It follows that only the terms proportional to
$e^{ i m \Omega t - i \phi(t)}$  can produce resonant driving.
The other non-resonant terms $\propto e^{ i m \Omega t + i \phi(t)}$
will produce corrections to the late time phase
evolution that are at least a factor of $\epsilon$ smaller than
those due to the resonant terms, and can be safely dropped.
Thus the resonant orbital frequency is
\be
\omega_0 = m \Omega + \omega_{\rm m}.
\label{eq:resonantfreq}
\ee
Using the formula (\ref{modefreq}) for the mode frequencies
$\omega_{\rm m} = \omega_{lm}$
we find for the two modes we consider
\bes
\label{eq:ourmodes}
\bea
\label{eq:ourmodes21}
m &=&1, \ \ \ \ \omega_{\rm m} = -\Omega/3, \ \ \ \ \ \omega_0 = 2\Omega/3, \\
m &=&2, \ \ \ \ \omega_{\rm m} = -2\Omega/3, \ \ \ \ \omega_0 =
4\Omega/3.
\label{eq:ourmodes22}
\eea
\ees

We next substitute the resonant terms from the overlap integrals
(\ref{forcingsII}) into Eq.\ (\ref{dotxrmode})
and use the formula
(\ref{eq:bans}) for $b$.
This finally gives
\bes\label{forcingsIII}
\bea
{\dot c}_{21} + i\omega_{21} c_{21} &=&\frac{i}{\omega_{21}} \sqrt{\frac{16 \pi}{5}}
\frac{I_{21} M^\prime R}{r^2} \omega (\omega - \Omega)  \nn \\
&& \times
\cos\left(\frac{3\psi}{2}\right)\cos\left(\frac{\psi}{2}\right) e^{i
  \Omega t-i \phi(t)}  \nn \\ \label{forcingsIII21} \\
{\dot c}_{22} + i\omega_{22} c_{22} &=&  \frac{i}{\omega_{22}}\sqrt{\frac{16 \pi}{5}}
\frac{I_{22} M^\prime R}{r^2} \omega (\omega - 2 \Omega)  \nn \\
&& \times
\sin\psi\cos^2\left(\frac{\psi}{2}\right) e^{2i
  \Omega t-i \phi(t)}. \nn \\
\label{forcingsIII22}
\eea
\ees
Note that for the aligned case $\psi=0$, only the $l=2,m=1$
mode is excited.  No modes are resonantly excited in
the anti-aligned case $\psi=\pi$.

The mode amplitude evolution equations (\ref{forcingsIII}) are exactly
of the form (\ref{nicedotx3}), with $m' = 1$. We can therefore
directly use the results of section \ref{sec:modeevolution} to obtain
the resonant response of each mode. From Eq.\ (\ref{fullxres1}) the
quantities needed to characterize this resonant response are the
resonant timescale $t_{\text{res}}$ [Eq.\ (\ref{tresdef})] and the
amplitude $A_{0,j}$ of the forcing term at resonance.  From
Eqs.\ (\ref{forcingsIII}) and (\ref{eq:ourmodes}) we obtain for the
values of these parameters
\bes\label{eq:resonantresponse}
\bea
t_{\text{res},21} &=& [96 \mathcal{M}^{5/3} (2\Omega / 3)^{11/3} / 5]^{-1/2}, \label{eq:tres21}\\
A_{0, 21} &=& \sqrt{\frac{16\pi}{5}}\frac{I_{21} M^\prime R}{r_0^2}\frac{2\Omega}{3}\cos\left(\frac{3\psi}{2}\right)\cos\left(\frac{\psi}{2}\right),\nn \\ && \label{eq:A021} \\
t_{\text{res},22} &=& [96 \mathcal{M}^{5/3} (4\Omega / 3)^{11/3} / 5]^{-1/2}, \label{eq:tres22}\\
A_{0, 22} &=& \sqrt{\frac{16\pi}{5}}\frac{I_{22} M^\prime R}{r_0^2}\frac{4\Omega}{3}\sin\psi\cos^2\left(\frac{\psi}{2}\right). \label{eq:A022}
\eea
\ees

\subsection{Effect of current quadrupole on orbital motion}
\label{sec:backreaction}

Since {\it r}-modes do not generate mass multipole moments to linear
order in the Lagrangian fluid displacement\footnote{This is correct
only to leading order in the star's spin frequency. There are
corrections to the leading order mode eigenfunction
(\protect{\ref{eq:rmodeeigenfunction}}) that scale as
$\Omega^2 R^3/M$.  As discussed in the introduction, these correction terms
can couple to the Newtonian tidal field and induce mass multipole
moments to linear order in the Lagrangian fluid displacement
\protect{\cite{1999MNRAS.308..153H}}.}, the
Newtonian equation of
motion for the star's center of mass worldline (and its companion's as
well) remain unchanged when such modes are driven. The leading order
correction to the equations of motion is a post-1-Newtonian tidal
interaction term involving the current quadrupole moment induced in
the star by the {\it r}-mode. The equations of motion including the
effect of this current quadrupole were derived in Ref.\ \cite{paperI}.
We specialize Eq.\ (6.11) of Ref.\ \cite{paperI} to two bodies with
masses $M_1=M$, $M_2
= M^\prime$, positions $\bm{z}_1(t)$ and $\bm{z}_2(t)$,
current quadrupoles $S^1_{ij}(t) = S_{ij}(t)$ and $S^2_{ij}(t)=0$, and
with all other mass and current moments equal to zero.  This gives
\bes\label{eomppcq}
\bea
\ddot{z}^i_1 &=& M^\prime  \frac{z_i}{r^3} - \frac{M^\prime}{M_{\rm
    t}} a^i \\
\ddot{z}^i_2 &=& -M \frac{z_i}{r^3} + \frac{M}{M_{\rm t}} a^i,
\eea
\ees
where $\bm{z} = \bm{z}_2 - \bm{z}_1$, $\bm{v} = \dot{\bm{z}}$, and
the contribution $\bm{a}$ to the relative acceleration $\ddot{\bm{z}}$
due to the current quadrupole is
\bea
\label{eq:relacc}
a^i &=&  - 4 \frac{M_{\rm t}}{M}
\left[\epsilon_{ipq}\dot{S}_{qr} \frac{z_p z_r}{r^5} -
  5\epsilon_{jpq} v_{p} S_{qr}\frac{z_{<i}z_jz_{r>}}{r^7} \right. \nn \\
&& \left. -
  5\epsilon_{ipq}v_jS_{qr}\frac{z_{<j}z_pz_{r>}}{r^7}\right].
\eea
Here the angular brackets denote taking the symmetric trace-free part
of the corresponding tensor, thus
\be
z_{<i}z_jz_{r>} = z_i z_j z_r - \frac{r^2}{5} ( \delta_{ij} z_r +
\delta_{ir} z_j + \delta_{jr} z_i ).
\ee

In Sec.\ \ref{sec:prelim} we showed that only the tangential component
$a_{\hat \phi}$ of the relative acceleration is needed to compute the
leading order phase lag parameter in the circular approximation.  That
tangential component is given by
\bea
\label{aphicq}
a_{\hat \phi} &=& \frac{1}{r \omega} \bm{v} \cdot \bm{a} \nn \\
&=&- \frac{4 M_{\rm t}}{M \omega r^6} (\bm{v} \times
\bm{z})_q\dot{S}_{qr}z_r.
\eea
Now for an axial fluid displacement $\bm{\xi}(t,\bar{\bm{x}})$, the induced current
quadrupole moment to linear order in $\bm{\xi}$ is given by the following
integral in \textit{inertial} coordinates
\bea\label{cq}
S_{qr} &=& \int \rho(r) \,\Big[-\Omega_k \xi_k x_{<q}x_{r>} + r^2
  \Omega_{<q}\xi_{r>}
  \nn \\ &&
  - 2\Omega_kx_k \xi_{<q}x_{r>} +
  x_{<q}\epsilon_{r>mn} x_m\dot{\xi}_n \Big] d^3x.\ \ \
\eea
Using the mode expansion (\ref{eq:modeexpansion}) and the mode
function (\ref{eq:rmodeeigenfunction}) we evaluate the integral
(\ref{cq}) for an $l=2$ {\it r}-mode.  Only the term proportional to
$\dot{\bm{\xi}}$ survives, and we obtain
\be\label{cqII}
S_{qr} = \frac{8\pi}{5}\sqrt{\frac{2}{3}}I_{2m} M R^3 \omega_{\rm m}  \,\,
\Re\Big\{e^{-im\Omega t} c_{2m}(t) \mathcal{Y}_{qr}^{2m}\Big\}.
\ee

Substituting the current quadrupole (\ref{cqII}) into the acceleration
(\ref{aphicq}) and using the orbit parameterization (\ref{eq:orbitp}) gives for the $m=1$ case
\bea
\label{eq:ahatphi21}
a_{{\hat \phi};21} &=&
\frac{4}{3}\sqrt{\frac{\pi}{5}}I_{21}\frac{\Omega}{r^3}M_{\rm t} R^3
\nn \\
&& \times \bigg\{ \cos(3\psi/2)\cos(\psi/2) \,  \left[ e^{i \phi - i \Omega t} ({\dot
     c}_{21} - i \Omega c_{21} ) \right] \nn \\
&&-\sin(3\psi/2)\sin(\psi/2) \, \left[ e^{-i \phi - i \Omega t} ({\dot
     c}_{21} - i \Omega c_{21}) \right] \bigg\} \nn \\ && + \text{c.c.}
\eea
Next we use the fact that in the vicinity of the resonance we have
${\dot c}_j = - i \omega_j c_j [1 + O(\epsilon)]$, from Sec.\
\ref{sec:modeev}, to eliminate the time derivative terms.
We also drop the terms proportional to $\sin(3\psi/2)\sin(\psi/2)$ as
they cannot give rise to secular contribution to the phase shift.
This yields
\bea
\label{eq:ahatphi21a}
a_{{\hat \phi};21} &=&
\frac{4}{3}\sqrt{\frac{\pi}{5}}I_{21}\frac{\Omega(\omega_{21} + \Omega)}{r^3}M_{\rm t} R^3
\cos(3\psi/2)\cos(\psi/2) \
\nn \\
&& \times \left[ - i e^{i \phi(t) - i \Omega t} c_{21}(t)  \right] + \,\text{c.c.}
\eea
Similarly we find for the $m=2$ mode
\bea
\label{eq:ahatphi22a}
a_{{\hat \phi};22} &=&
\frac{8}{3}\sqrt{\frac{\pi}{5}}I_{22}\frac{\Omega(\omega_{22} + 2\Omega)}{r^3}M_{\rm t} R^3
\sin\psi\cos^2(\psi/2) \
\nn \\
&& \times
\left[ - i e^{i \phi(t) - 2 i \Omega t} c_{22}(t)  \right] + \, \text{c.c.}
\eea
These tangential accelerations
(\ref{eq:ahatphi21a}) and (\ref{eq:ahatphi22a})
coincide with the prediction (\ref{eq:ahatphi1}) of the Newtonian
model of Sec.\ \ref{sec:prelim}, when we use the formulae (\ref{eq:A021}) and
(\ref{eq:A022}) for the amplitudes at resonance $A_{0,j}$ of the mode
forcing terms.  We can therefore use the results of this Newtonian
model.  Substituting the amplitudes and resonance timescales
(\ref{eq:resonantresponse}) into the formula (\ref{answers}) for the phase shift
$\Delta \phi$ and using Eqs.\ (\ref{eq:bans}) and (\ref{eq:ourmodes})
finally yields
\bes
\bea
\label{eq:deltaphi21ans}
\Delta \phi_{21} &=&
-\frac{5\pi^2}{192}\left(\frac{2}{3}\right)^{2/
3}I_{21}^2
\cos^2\left(\frac{3\psi}{2}\right)\cos^2\left(\frac{\psi}{2}\right)
\nn \\
&& \times \frac{\Omega^{2/3} R^4}{M^\prime M^2 M_{\rm t}^{1/3}}, \\
\Delta \phi_{22} &=& -\frac{5 \pi^2}{192}
\left(\frac{4}{3}\right)^{2/3}
I_{22}^2 \sin^2\psi\cos^4\left(\frac{\psi}{2}\right) \nn \\
&& \times \frac{\Omega^{2/3} R^4}{M^\prime M^2 M_{\rm t}^{1/3}}.
\label{eq:deltaphi22ans}
\eea
\ees

\subsection{Numerical values for specific neutron star models}
\label{sec:evaluate}

\subsubsection{Barotropic stars}

In the case of barotropic stars (i.e.\ stars without buoyancy forces) the formula
(\ref{eq:rmodeeigenfunction}) for
the mode function only applies for $|m|=l$, in which case $f_{lm}(r)
\propto r^l$.  In this case, using the normalization condition (\ref{modenorm1}),
the formula (\ref{eq:radialint}) for the parameter $I_{22}$ evaluates to
\be
I_{22} = \left[ \frac{1}{M R^4} \int_0^R  \rho(r) r^6 dr
\right]^{1/2}.
\ee
This gives
\be
I_{22} = \sqrt{\frac{3}{28\pi}} \simeq 0.185
\ee
for constant density stars, while for a $\Gamma = 2$ polytrope
for which $\rho(r) = M \sin(\pi r/R) / (4 R^2 r)$
we obtain
\be
I_{22} = \frac{\sqrt{120 - 20\pi^2 + \pi^4}}{2 \pi^{5/2}} \simeq 0.128.
\label{eq:I22p}
\ee
For other stable polytropes, one needs to solve the Lane-Emden equation numerically to obtain the number $I_{22}$. For $\Gamma = 4/3, 3/2$ and 3 respectively, the number $I_{22}$ turns out to be equal to $0.0437$, $0.0812$ and $0.155$, a variation of about a factor of 5 over the range of all polytropic indices. Evaluating the phase shift (\ref{eq:deltaphi22ans}) for the equal mass
case $M=M^\prime$ and at the value $\psi = \pi/3$ of the inclination
angle which maximizes the mode driving gives
\be
\Delta \phi_{22} = - 0.05 \ R_{10}^4 f_{s100}^{2/3} M_{1.4}^{-10/3}
\label{eq:ans2ssa0}
\ee
for the constant density case and
\be
\Delta \phi_{22} = - 0.03 \ R_{10}^4 f_{s100}^{2/3} M_{1.4}^{-10/3}
\label{eq:ans2ssa}
\ee
for the $\Gamma=2$ case.  Here we have defined $R = 10
R_{10} \, {\rm km}$, $M = 1.4 M_{1.4} M_\odot$, and $\Omega/(2\pi) =
100 f_{s100} \, {\rm Hz}$.
As discussed in the introduction, these phase shifts may be marginally
detectable by gravitational wave interferometers for nearby inspirals.

Consider next the case $m \ne l$.  As discussed above there are no
purely axial modes of this type in barotropic stars.  There are modes
analogous to the $r$-modes which have both axial and polar pieces; these
are the inertial or hybrid modes \cite{1999ApJ...521..764L}.
These modes are somewhat difficult to compute in barotropic stars,
but are easier to compute in stars with buoyancy where they are
purely axial.  Therefore for the $m\ne l$ case we will switch to using
a slightly more realistic neutron star model which includes buoyancy.

\subsubsection{Stars with buoyancy}

In stars with buoyancy, the formula (\ref{eq:rmodeeigenfunction}) for
the mode functions is valid for all $m$ and $l$.  However in this case
the radial mode
function $f_{lm}(r)$ is not a simple power law, but instead must be
obtained from solving numerically a complicated Sturm-Liouville problem
\cite{1981A&A....94..126P}.
The stellar model we use consists of a $\Gamma = 2$ polytrope
for the background star, with the perturbations characterized by a
constant adiabatic index $\Gamma_1 \ne \Gamma$ chosen so that
the Brunt-V\"{a}is\"{a}l\"{a} frequency evaluated at a location halfway
between the center of the star and its surface is 100 Hz.
See Refs.\
\cite{1987MNRAS.227..265F,1992ApJ...395..240R,1994MNRAS.270..611L} for
detailed models of the buoyancy force and corresponding $g$-modes of
which our chosen model is representative.
Our results below are fairly robust with respect to the choice of
buoyancy model, varying by no more than $\sim 20\%$ for a $\sim 50\%$
change in the Brunt-V\"{a}is\"{a}l\"{a} frequency evaluated at a location halfway
between the center of the star and its surface.

We compute the radial mode function $f_{lm}$ for modes with different
numbers $n$ of radial nodes.  The net phase shift is obtained by
summing over all these modes, so we use an effective value of
$I_{21}^2$ obtained by summing over $n$.  We find that it is sufficient to
sum over $n=0,1$ and $2$ in order to obtain two digits of accuracy in
$I_{21}^2$, and we obtain
\be
I_{21}^2 = 0.016.
\ee
We substitute this into Eq.\ (\ref{eq:deltaphi21ans}) and
evaluate the phase shift for the equal mass
case, and at the value $\psi = 0$ of the inclination
angle which maximizes the mode driving.  This gives
\be
\Delta \phi_{21} = - 0.03 \ R_{10}^4 f_{s100}^{2/3} M_{1.4}^{-10/3}.
\label{eq:ans21ssa}
\ee

\section{Conclusions}

In this paper we have studied in detail the gravitomagnetic resonant
tidal excitation of {\it r}-modes in coalescing neutron star
binaries.  We first analyzed the simpler case of Newtonian resonant
tidal driving of normal modes. We showed, by integrating the equations
of motion of the mode-orbit system, that the effect of the resonance
to leading order is an instantaneous shift in the
orbital frequency of the binary at resonance. The effect of the modes
on the orbital motion away from resonance is a small correction to this
leading order effect. We then showed that for the case of
gravitomagnetic tidal driving of {\it
  r}-modes, the equations of motion for the mode-orbit system can be
manipulated into a form analogous to the Newtonian case. We then made
use of the Newtonian results to compute the instantaneous change in
frequency due to the driving of the {\it r}-modes.
This shift in frequency is negative, corresponding to energy being
removed from the star and added to the orbital motion.  This sign of
energy transfer occurs only for modes which satisfy the
Chandrasekhar-Friedman-Schutz instability criterion.
We also argued that the resonances are at best marginally detectable
with LIGO.

\begin{acknowledgments}
We thank Marc Favata and Dong Lai for helpful discussions and for
comments on the manuscript.
This research was supported by the Radcliffe Institute and by NSF
grants PHY-0140209 and PHY-0457200.  \'ER was supported by
NATEQ (Fonds qu\'{e}b\'{e}cois de la recherche sur la nature et les
technologies), formerly FCAR.
\end{acknowledgments}

\appendix

\section{Computation of the gravitomagnetic tidal tensor}
\label{app:Y}

In this appendix we compute the gravitomagnetic tidal
tensor ${\cal B}_{ij}(t)$ using two
different methods, first using the global inertial frame, and second
directly in the star's local asymptotic rest frame.

We denote by $({\tilde t},{\tilde x}^i)$ the global harmonic
center-of-mass frame.  In this frame the Newtonian potential and
gravitomagnetic potential are
\bes
\label{eq:potin}
\bea
\Phi(\tilde{t},\tilde{\bm{x}}) &=& - \frac{M}{ | \tilde{\bm{x}}
-\bm{z}_1(\tilde{t}) | } - \frac{M^\prime}{|\tilde{\bm{x}} -
\bm{z}_2(\tilde{t})|},\\
\bm{\zeta}(\tilde{t},\tilde{\bm{x}}) &=& - \frac{4 M
\dot{\bm{z}}_1(\tilde{t}) }{ | \tilde{\bm{x}}
-\bm{z}_1(\tilde{t}) | } - \frac{4 M^\prime
\dot{\bm{z}}_2(\tilde{t}) }{|\tilde{\bm{x}} -
\bm{z}_2(\tilde{t})|}.
\eea
\ees
The post-Newtonian scalar potential $\psi$ will not be needed in what follows.
We now transform to a body-adapted coordinate system $(t,x^i)$ of
the star of mass $M$.  We use the general coordinate transformation
from one harmonic, conformally Cartesian coordinate system to another
given by Eq.\ (2.17) of Ref.\ \cite{paperI} with $(t,x^i)$ replaced
by $(\tilde{t},\tilde{x}^i)$ and $(\bar{t},\bar{x}^i)$ replaced by $(t,x^i)$.
That coordinate transformation is parameterized by a number of free
functions: a function $\alpha_{\rm c}(\tilde{t})$ governing the normalization
of the time coordinate at Newtonian order; a function $d^i(\tilde{t})$
governing the translational freedom at Newtonian order (denoted $z^i$
in Ref.\ \cite{paperI}); a function $R_k(\tilde{t})$ governing the
angular
velocity of the coordinate system; a function $h^i_{\rm c}(\tilde{t})$
governing the translational freedom at post-Newtonian order, and a
free harmonic function $\beta_{\rm c}(\tilde{t},\tilde{x}^i)$.
The first four of these functions are fixed by the requirements listed
at the start of Sec.\ \ref{sec:rmodeoverview} above.  The fifth
function $\beta_{\rm c}$ parameterizes the remaining gauge freedom,
under which the tidal tensors are invariant.
In fact the gravitomagnetic tidal tensors do not depend on $h^i_{\rm
c}$, so we can use a coordinate transformation with $h^i_{\rm c} =
\beta_{\rm c} =0$.  The required values of the other functions
are given by
\bes
\bea
\dot{\alpha}_{\rm c}(\tilde{t}) &=& \frac{1}{2}
\dot{\bm{z}}_1^2(\tilde{t}) -\Phi_{\rm ext}[\tilde{t},\bm{z}_1(\tilde{t})], \\
\bm{d}(\tilde{t}) &=& \bm{z}_1(\tilde{t}), \\
\dot{\bm{R}}(\tilde{t}) &=& \frac{1}{2} \bm{\nabla} \times
\bm{\zeta}_{\rm ext}[\tilde{t},\bm{z}_1(\tilde{t})] + 2
\dot{\bm{z}}_1(\tilde{t}) \times \bm{\nabla} \Phi_{\rm
ext}[\tilde{t},\bm{z}_1(\tilde{t})] \nn \\
&& +\frac{1}{2} \ddot{\bm{z}}_1(\tilde{t}) \times
\dot{\bm{z}}_1(\tilde{t}),
\eea
\ees
where $\Phi_{\rm ext}$ and $\bm{\zeta}_{\rm ext}$ denote the second
terms on the right hand sides of Eqs.\ (\ref{eq:potin}).

Using this coordinate transformation we can compute the transformed
gravitomagnetic potential $\bm{\zeta}(t,\bm{x})$ in the body frame $(t,x^i)$.
We can then extract the gravitomagnetic tidal moment ${\cal B}_{ij}(t) =
- H_{ij}(t)/2$ by comparing with the general multipolar expansion of
$\bm{\zeta}$
given in Eq.\ (3.5c) of Ref.\ \cite{paperI}.  That extraction method is
equivalent to
throwing away the pieces of $\bm{\zeta}$ that diverge at the origin
$\bm{x}=0$ and evaluating ${\cal B}_{ij} = - B_{(i,j)}$ at $\bm{x}=0$,
where $\bm{B} = \bm{\nabla} \times \bm{\zeta}$ is the gravitomagnetic field.
After eliminating the acceleration terms
using the Newtonian equations of motion
the result is the expression (\ref{eq:calBans}).

An alternative, simpler method of computation is to solve the
post-1-Newtonian field equations directly in the body frame.  The only
complication here is that the boundary condition on the potentials
$\Phi$ and $\bm{\zeta}$ as $r \to \infty$ are no longer $\Phi \to 0$,
$\bm{\zeta} \to 0$, since the body frame is accelerated and rotating
with respect to the global inertial frame.  In the body frame the two
stars are located at $\bm{x} =0$ and at $\bm{x} = \bm{z}(t) \equiv
\bm{z}_2(t) - {\bm z}_1(t)$.  The solutions to the field
equations are thus
\bes
\label{eq:potbody}
\bea
\Phi(t,\bm{x}) &=& - \frac{M}{ | \bm{x} | } - \frac{M^\prime}{|\bm{x} -
\bm{z}(t)|} + \Phi_{\rm bc}(t,\bm{x}),\\
\bm{\zeta}(t,\bm{x}) &=&
- \frac{4 M^\prime
\dot{\bm{z}}(t) }{|\bm{x} -
\bm{z}(t)|} + \bm{\zeta}_{\rm bc}(t,\bm{x}).
\eea
\ees
Here $\Phi_{\rm bc}$ and $\bm{\zeta}_{\rm bc}$ are solutions of
the Laplace equation that are determined by the large-$r$ boundary
conditions.   They are determined by the functions $\alpha_{\rm c}$,
$d^i$, $h^i_{\rm c}$ and $R^k$ discussed above.  The key point is that
these terms do not contribute to the gravitomagnetic tidal tensors;
this can be seen from the explicit form of these terms given in Eq.\
(3.38c) of Ref.\ \cite{paperI}.  Therefore we can drop the terms
$\Phi_{\rm bc}$ and $\bm{\zeta}_{\rm bc}$ and extract the
gravitomagnetic tidal tensor ${\cal B}_{ij}(t)$ using the method
discussed above; the result is again given by Eq.\ (\ref{eq:calBans}).

\section{Alternative derivation of time delay parameter using
conservation of energy}
\label{sec:energycons}

In this appendix we give an alternative derivation of the time delay
parameter $\Delta t$ using conservation of energy, for the
Newtonian model of Secs.\ \ref{sec:prelim}, \ref{sec:modeevolution} and
\ref{sec:phasing} and also for the $r$-mode driving of Sec.\
\ref{sec:rmodedriving}.  The energy conservation method is simpler
to use than the method of integration of the orbital equations of
motion used in the body of the paper.  However, it only gives
information about the time delay parameter $\Delta t$, and says
nothing about the phase shift parameter $\Delta \phi$.  In order to
derive the relation (\ref{eq:qq}) between these two parameters, an analysis
of the orbital equations of motion is necessary.

\subsection{Newtonian mode driving}

Consider the evolution of the binary from some initial frequency
$\omega_i$ below resonance to some final resonance $\omega_f$ after
resonance.  We denote by
\be
{\dot \omega}(\omega) = {\dot \omega}_{\rm pp}(\omega) + \delta {\dot
  \omega}(\omega)
\ee
the time derivative of the orbital frequency $\omega$ as a function of
$\omega$; this consists of a sum of the inspiral rate ${\dot \omega}_{\rm
  pp}(\omega)$ for point particles [Eq.\ (\ref{eq:inspiral0}) above] together with
a perturbation $\delta {\dot \omega}(\omega)$ due to the mode coupling.
The total time taken for the binary to evolve from $\omega_i$ to
$\omega_f$ can be written as
\bea
t(\omega_i \to \omega_f) &=& \int_{\omega_i}^{\omega_f} \frac{d\omega
}{\dot \omega(\omega)} \nn \\
&=& \int_{\omega_i}^{\omega_f} \frac{d\omega
}{\dot \omega_{\rm pp}(\omega)} - \int_{\omega_i}^{\omega_f}
\frac{\delta {\dot \omega}(\omega)}{\dot \omega_{\rm pp}(\omega)^2}
d\omega. \ \ \ \
\label{eq:timepp}
\eea
Here on the second line we have linearized in $\delta {\dot \omega}$.
The first term in Eq.\ (\ref{eq:timepp}) is the time it would take for
point particles, and by comparing with Eq.\ (\ref{eq:phiapproxprime})
we can identify the second term with the negative of the time delay
parameter $\Delta t$:
\be
\Delta t = \int_{\omega_i}^{\omega_f}
\frac{\delta {\dot \omega}(\omega)}{\dot \omega_{\rm pp}(\omega)^2}
d\omega.
\label{eq:Deltatintegral}
\ee

Consider next the total energy radiated into gravitational waves
between $\omega_i$ and $\omega_f$.  This radiated energy can be
written as
\bea
\label{eq:energypp0}
E_{\rm rad}(\omega_i \to \omega_f) &=& \int_{\omega_i}^{\omega_f}
\frac{{\dot E}_{\rm gw}(\omega)
}{\dot \omega(\omega)} d\omega \\
&=& \int_{\omega_i}^{\omega_f} \left[\frac{{\dot E}_{\rm gw}(\omega)
}{\dot \omega_{\rm pp}(\omega)} -
\frac{{\dot E}_{\rm gw}(\omega) \delta {\dot \omega}(\omega)}{\dot
  \omega_{\rm pp}(\omega)^2}\right]
d\omega,\nn \\
&&
\label{eq:energypp}
\eea
where ${\dot E}_{\rm gw}(\omega)$ is the gravitational wave luminosity
and as before we have linearized in $\delta {\dot \omega}(\omega)$.
A key point now is that the gravitational wave luminosity ${\dot
  E}_{\rm gw}(\omega)$ is to a good approximation unaffected by the
mode coupling, so that it is the same function for the point-particle
inspiral and for the true inspiral.\footnote{This approximation
is valid for computing the resonance phase shift.  It is not valid, however, for
computing the much smaller orbital phase correction in the adiabatic regime before
resonance.  In that regime one must include the gravitational
radiation from the mode, which is phase
coherent with that from the orbit
\protect{\cite{1993ApJ...406L..63L,Hinderer:2005}}.}
Therefore the first term in Eq.\
(\ref{eq:energypp}) is just the energy radiated between $\omega_i$ and
$\omega_f$ for a point-particle inspiral, namely
$$
E_{\rm orb}(\omega_i) - E_{\rm orb}(\omega_f).
$$
Here $E_{\rm orb}(\omega)$ is
the orbital energy of the binary at frequency $\omega$.

Consider now the evaluation of the second term in Eq.\
(\ref{eq:energypp}).  The function $\delta {\dot \omega}(\omega)$ is
peaked around the resonance $\omega = \omega_0$ with width $\Delta
\omega \sim {\dot \omega}(\omega_0) t_{\rm res} \sim \epsilon
\omega_0$.  The fractional variation in the luminosity ${\dot E}_{\rm
  gw}(\omega)$ over this frequency band is of order $\sim \epsilon$,
so to a good approximation we can replace ${\dot E}_{\rm gw}(\omega)$
with ${\dot E}_{\rm gw}(\omega_0)$ and pull it outside the
integral.\footnote{Note that this approximation can only be applied
to the second term in Eq.\ (\ref{eq:energypp}) and not to the entire
integral (\ref{eq:energypp0}); the corresponding errors in energy
would in that case be larger than the energy we are computing here.}
The remaining integral is just the formula (\ref{eq:Deltatintegral})
for the time delay parameter $\Delta t$, and we obtain
\bea
E_{\rm rad}(\omega_i \to \omega_f) &=&
E_{\rm orb}(\omega_i) - E_{\rm orb}(\omega_f) \nn \\
&&- {\dot E}_{\rm  gw}(\omega_0) \Delta t \left[ 1 + O(\epsilon)\right].
\label{eq:Eradformula}
\eea
Finally global conservation of energy gives
\bea
E_{\rm orb}(\omega_i) + E_{\rm star}(\omega_i) &=& E_{\rm rad}(\omega_i \to \omega_f)+ E_{\rm
  orb}(\omega_f) \nn \\
&&  + E_{\rm star}(\omega_f),
\eea
where $E_{\rm star}$ is the total energy of the star.\footnote{We
neglect here the tidal interaction energy corresponding to the last two
terms in the Hamiltonian (\ref{eq:Hamiltonian}).
Although this
energy is comparable to the mode energy during resonance, at late
times after resonance it is smaller than the mode energy by a factor
$\sim \epsilon$.}  Defining
\be
\Delta E_{\rm star} = E_{\rm star}(\omega_f) - E_{\rm star}(\omega_i)
\ee
and combining this with Eq.\ (\ref{eq:Eradformula})
yields
\be
\Delta E_{\rm star} = {\dot E}_{\rm gw}(\omega_0) \Delta t \left[ 1 +
  O(\epsilon) \right].
\label{eq:ec1}
\ee

Note that a simple interpretation of the formula
(\ref{eq:Eradformula}) is that the resonance can be
treated as an instantaneous change of frequency of magnitude
$\Delta \omega = {\dot \omega}_{\rm pp}(\omega_0) \Delta t$ and corresponding
instantaneous change in orbital energy of ${\dot E}_{\rm gw}(\omega_0)
\Delta t$.  However this interpretation is not completely valid since
the resonance is not instantaneous: the energy radiated during
resonance is larger than the energy $E_{\rm mode}$ absorbed by the
mode by a factor $\sim 1/\epsilon$.  The formula
(\ref{eq:Eradformula}) must therefore be derived by integrating over
the resonance as we have done above.

The change in the energy of the star
$\Delta E_{\rm star}$ consists of the rotating-frame energy in the mode
$E_{\rm mode}$, the remaining portion of the energy in the mode due to
its angular momentum, and the change in the rotational kinetic energy of the
star.  The total change
can be computed from a prescription detailed in Appendix K of Ref.\ \protect{\cite{Schenk:2001zm}}, which gives
\footnote{Note that the splitting of the total energy of a perturbed
  star into mode and rotational contributions depends on the choice of
  a background or reference uniformly-rotating star.  In the formalism
  of Ref.\ \cite{Schenk:2001zm}, such a choice is determined by the
  use of Lagrangian displacement as the fundamental variable.  Other
  choices such as Eulerian velocity perturbation yield different
  results for this splitting.  The same issue applies to angular
  momentum, and explains why the angular momentum of $r$-modes was
  found to be zero in Ref.\ \cite{Levin:1999jf}, in disagreement with the results of
  Ref.\ \cite{Schenk:2001zm}.}
\bea
\Delta E_{\rm star} &=& \frac{E_{\rm mode}}{\omega_j}[\omega_j + m_j\Omega].
\label{eq:modeenergy}
\eea
Here $E_{\rm mode}$ is given by $E_{\rm mode} = b_j \omega_j |c_j|^2 =
b_j \omega_j B_l^2$, cf.\ Eq.\ (\ref{eq:Hamiltonian}) above.
Combining this with Eq.\ (\ref{eq:ec1}) gives a formula for $\Delta t$
which agrees with our earlier formula (\ref{eq:Deltatans}). Equation (\ref{eq:modeenergy}) is computed as follows. Equation (K17) of \cite{Schenk:2001zm} gives the formula for the physical energy due to the fluid perturbation. The first term of (K17) is the rotating-frame energy of the mode and the two other terms are angular momentum contributions to the total energy of the perturbation. The rotating-frame energy is given by (K21) and the angular momentum contribution is computed using the binary's orbital equations of motion including mode coupling, along with conservation of total angular momentum \footnote{Over the resonance timescale, the amount of angular momentum radiated away by gravitational waves is negligible for this computation.}, following the work of Ref.\ \cite{RPA:2006}.

We conclude that using energy conservation gives a simple method to
evaluate the time delay parameter $\Delta t$, without computing the
evolution of the orbital variables.
However, energy conservation does not give any direct information
about the phase shift parameter $\Delta \phi$; for that one must
analyze the resonance itself.  As we have discussed, the result of
this analysis is the relation (\ref{eq:qq}) between $\Delta \phi$ and
$\Delta t$, and using this relation together with Eq.\ (\ref{eq:ec1}) we obtain
a formula for the phase shift in terms of the mode energy:
\be
\Delta \phi = \left(\frac{m_j\omega_0}{\omega_j}\right)\frac{\omega_0 E_{\rm mode} }{ {\dot E}_{\rm
    gw}(\omega_0)}.
\ee
In Ref.\ \cite{1999MNRAS.308..153H} energy conservation was used to compute the phase shift, although there $\Delta
\phi$ was misinterpreted as the asymptotic value of the phase
perturbation after the resonance.  The fact that the phase
perturbation grows at late times [cf.\ Eq.\ (\ref{eq:signature1}) above]
was noted in Refs.\ \cite{1994ApJ...426..688R,Sharon}.

\subsection{Gravitomagnetic mode driving}

The argument of the previous subsection can be carried over to
post-1-Newtonian gravity with a few minor modifications.
First, the total conserved post-1-Newtonian energy $E$ can be split up
into two pieces, an ``orbital'' piece $E_{\rm orb}$ and a remaining
piece, as follows.  On a given constant time slice we can compute the
post-1-Newtonian positions and velocities of the stars'
center-of-mass worldlines, and also the post-1-Newtonian masses of the
bodies \cite{1991PhRvD..43.3273D,paperI}.
The post-1-Newtonian masses will evolve with time, so we choose the values
of these masses at $t \to -\infty$ when the stars are far apart.
We then compute the total post-1-Newtonian energy for a system of
point particles with the same positions, velocities and (constant)
masses; this defines the orbital energy $E_{\rm orb}$ of the
system. The remaining piece of the total energy consists of the mode
excitation energy $E_{\rm mode}$, together with an interaction energy
$E_{\rm int} \propto S_{ij} {\cal{B}}_{ij}$ which we can neglect for the
same reason as in the Newtonian case.\footnote{At late times it is
smaller than $E_{\rm mode}$ by a factor of $\epsilon$.}
With these definitions the arguments of the last subsection are valid
for gravitomagnetic driving.

In the Newtonian case it was most convenient to evaluate the energy
$E_{\rm mode}$ transferred into the stellar modes by using the
explicit formula (\ref{eq:modeenergy}) for the mode energy.
In the post-Newtonian case it is more convenient to use a different
technique.
A formula for total tidal work done on the star by the gravitomagnetic
driving can be derived by surface integral techniques
\cite{Zhang:1985qz}; the result is
\be
\frac{d E_{\rm mode}}{dt} = -\frac{2}{3} {\cal B}_{ij}(t) {\dot S}_{ij}(t).
\ee
See Refs.\
\cite{Thorne:1998kt,Purdue:1999gk,Booth:2000ka,Favata:2000vn} for
further discussion of the derivation method and of similar tidal work
formulae.

We now use the explicit formula (\ref{eq:calBans}) for the
gravitomagnetic tidal
moment ${\cal B}_{ij}$, and compare the resulting expression with the
formula (\ref{aphicq}) for the tangential acceleration $a_{\hat
\phi}$ due to the current quadrupole moment, specializing to a
circular orbit.  The result is
\be
\frac{d E_{\rm mode}}{d t} = \mu \omega r a_{\hat \phi}.
\ee
Using the formula (\ref{eq:ec1}) for $\Delta t$ and integrating over the
resonance gives
\be
\Delta t = \frac{\mu \omega_0 r_0}{{\dot E}_{\rm gw}(\omega_0)} \,
\int dt \, a_{\hat \phi}(t).
\ee
This is the same result as was obtained in
Secs.\ \ref{sec:evol2} and \ref{sec:backreaction} for the time
delay parameter; those sections also involved integrating $a_{\hat
\phi}$ with respect to time, and it
can be checked that the prefactors match.
Thus energy conservation gives the same result as the method
of direct integration of the equations of motion in the
gravitomagnetic case.

\newcommand{\apjl}{Astrophys. J. Lett.}
\newcommand{\aap}{Astron. and Astrophys.}
\newcommand{\cmp}{Commun. Math. Phys.}
\newcommand{\grg}{Gen. Rel. Grav.}
\newcommand{\lr}{Living Reviews in Relativity}
\newcommand{\mnras}{Mon. Not. Roy. Astr. Soc.}
\newcommand{\pr}{Phys. Rev.}
\newcommand{\prsl}{Proc. R. Soc. Lond. A}
\newcommand{\ptrsl}{Phil. Trans. Roy. Soc. London}

\bibliographystyle{prsty}
\bibliography{rmodetidal}

\begin{thebibliography}{10}
\bibitem{hcdef}
This signal strength $h_c(f)$ is defined so that the squared signal to noise
  ratio, averaged over the orientations of source and detector, is given by
  $\int d (\ln f) h_c(f)^2 / h_{\rm rms}(f)^2$.
\end{thebibliography}

\end{document}